\documentclass[acmlarge,nonacm]{acmart}

\usepackage{xcolor}

\AtBeginDocument{
  \color{black}
}

\AtBeginDocument{%
  }

\begin{document}

\title{``Can Computing Systems Eventually Disappear?'': A Long-Term Tracking Study on the Use of Diabetes Computing Systems by Patients with Diabetes}

\author{Yibo Meng}
\authornote{Both authors contributed equally to this research.}

\orcid{0009-0009-0370-5456}
\email{yibom2@illinois.edu}
\affiliation{%
  \institution{Cornell University}
  \city{New York}
  \state{New York State}
  \country{the United States}
}

\author{Jinyuan Zhan}
\authornotemark[1]
\orcid{0009-0003-7266-333X}
\affiliation{%
  \institution{National University of Singapore}
  \country{Singapore}
}

\author{Yao Lyu}
\orcid{0000-0003-3962-4868}
\affiliation{%
  \institution{University of Michigan}
  \city{Ann Arbor}
  \state{Michigan}
  \country{United States}
}

\author{Zhuoran Lu}
\authornote{Corresponding authors.}
\orcid{0000-0002-1079-2043}
\affiliation{%
  \institution{Purdue University}
  \city{West Lafayette}
  \state{Indiana}
  \country{United States}
}

\author{Chengxi Zang}
\authornotemark[2]
\orcid{0000-0002-8244-9551}
\affiliation{%
  \institution{Cornell University}
  \city{New York}
  \state{New York}
  \country{United States}
}

%%
%% By default, the full list of authors will be used in the page
%% headers. Often, this list is too long, and will overlap
%% other information printed in the page headers. This command allows
%% the author to define a more concise list
%% of authors' names for this purpose.

%%
%% The abstract is a short summary of the work to be presented in the
%% article.
\begin{abstract}
    Ubiquitous computing has long pursued Weiser's vision of technologies that ``disappear'' into everyday life. As wearable health systems advance, they increasingly substitute for rather than scaffold users' judgment, yet the long-term consequences remain underexplored. We report an eight-year, five-wave longitudinal interview study from 2017 to 2025 with 37 patients with diabetes, complemented by interviews with 15 clinicians, designers, and vendors. Across three technological generations including glucometers, continuous glucose monitoring, and predictive and automated insulin delivery systems, we identify a novel user state, continued use with cognitive disengagement. In this state, devices remain in use but active interpretation and questioning recede, leaving users to respond only to alarms. This state emerges from functional expansion, delegation of interpretive authority, and information fatigue, and is reinforced by clinical prioritization, information compression, and automation.It results in less frequent independent health judgment and a shift from understanding to responding.We argue that future diabetes computing systems should reduce unnecessary cognitive work while preserving users' opportunities to interpret, contextualize, and question system outputs.
\end{abstract}

%%
%% The code below is generated by the tool at http://dl.acm.org/ccs.cfm.
%% Please copy and paste the code instead of the example below.
%%


%%
%% Keywords. The author(s) should pick words that accurately describe
%% the work being presented. Separate the keywords with commas.

\received{1 August 2026}
\received[revised]{21 September 2026}
\received[accepted]{8 October 2026}

%%
%% This command processes the author and affiliation and title
%% information and builds the first part of the formatted document.
\maketitle

\section{Introduction}
Mark Weiser's foundational vision of ubiquitous computing imagined technologies that ``weave themselves into the fabric of everyday life until they are indistinguishable from it'' \cite{weiser1991computer,weiser1996calm}. In the domain of personal health, this vision has become the explicit design target of an entire generation of products. Diabetes self-management offers a revealing case of this vision: its computing systems are used in everyday life over long periods and have rapidly taken over more of the work once performed by users. Continuous glucose monitoring (CGM) has replaced intermittent finger-prick testing with continuous time-series sensing; predictive algorithms anticipate hypo- and hyperglycemic events before they happen; automated insulin delivery (AID) closes the loop between sensing and action with minimal user intervention. Each generation takes less effort than the previous one.

However, continued use may hide a gradual shift from interpreting data to relying on system judgments. Over years of routine use, this shift could become consequential when users must act without reliable system support. Prior work gives us reasons to take this problem seriously. Automation can induce complacency or skill decay \cite{parasuraman1997humans,parasuraman2010complacency}, how seamless tracking can bypass the reflection necessary for deep understanding \cite{li2010stagebased}, and how unexamined trust erodes data awareness \cite{barth2017privacyparadox}. 
Despite these findings, much of this work examines users at a single point in time or over short periods of use. It therefore remains unclear whether early reductions in engagement fade as users gain experience, remain stable, or become more pronounced as systems take over more work. Answering this question requires following the same users over time and across successive generations of technology to capture gradual shifts in user--system engagement through years of routine use. Accordingly, we report an eight-year (2017--2025), five-wave longitudinal study following the same cohort of people with diabetes across three generations of diabetes technology. The research questions guiding this study are:

\begin{itemize}
    \item RQ1: How do people with diabetes engage with and interpret data from digital diabetes-management technologies over long-term use as system capabilities evolve?
    \item RQ2: What design problems emerge as user's cognitive engagement changes over long-term usage?  
\end{itemize}

Our primary cohort consists of 37 patients with diabetes. The same interview protocol was applied at each wave, allowing us to trace within-subject change in usage context, interpretive strategy, and decision logic across three sociotechnical generations. To complement the user side, we conducted single-wave interviews in 2025 with 15 practitioners (5 clinicians, 5 designers, 5 vendors) involved in the design and deployment of these systems. The study aims not at population-level generalization but at process-level mechanism: how the user--system relationship evolves, what design conditions shape it, and what consequences accumulate.

Our analysis surfaces three recurring stages in how users engage with these systems: (1) exploration and sensemaking, in which users actively compare, verify, and build causal models of their data; (2) collaboration and delegation, in which users increasingly respond to system conclusions rather than independently interpreting the underlying data; and (3) cognitive disengagement, in which continued behavioral use decouples from self-initiated interpretation. We synthesize the endpoint of this account into a previously under-theorized user state—\emph{continued use with cognitive disengagement}. The design-side data reveal that this state is not a contingent failure but is \emph{pre-configured} by three rational, mutually reinforcing logics: clinical prioritization of timely intervention over patient understanding, designerly compression of complex data into ``conclusive signals,'' and an engineering imperative to minimize user operations. The consequences materialize in two layers: health risk, in which independent judgment of bodily state is replaced by default trust in system output; and a transfer of burden from understanding complex data to continuously responding to system prompts—a hidden cost that accumulates as fatigue and further withdrawal rather than as accumulated competence.

We make four contributions. First, empirically, we report what is, to our knowledge, the first eight-year, multi-wave, multi-generational longitudinal account of the same patient cohort across CGM, predictive systems, and AID. Second, conceptually, we name and theorize \emph{continued use with cognitive disengagement} as a third state that fills a gap left by abandonment, complacency, and deskilling research. Third, mechanistically, we link this user-side trajectory to a design-side confluence of clinical, designerly, and engineering logics, providing an integrated explanation that joins automation-trust theory, lived-informatics theory, and human--data interaction in a single longitudinal frame. Fourth, generatively, we argue for a redesign of ubiquitous health computing from a judgment substitute toward automation that preserves interpretive engagement, and we propose three design implications: providing layered explanations at consequential moments; automating routine work while preserving users' contextual contribution; and designing for failures in sensing, alerting, and perception.

Beyond diabetes care, the pattern we describe may have broader implications for the IMWUT community. Wherever ubiquitous computing systems aim to disappear into users' lives, whether in clinical AI, smart-home infrastructure, productivity assistants, or emerging agentic systems, the same three drivers (clinical/functional prioritization, informational compression, and engineering minimization of user operations) could plausibly be at work, though whether they produce a comparable pattern toward cognitive disengagement remains an open empirical question. Whether "disappearance" produces calm or cognitive impoverishment is therefore not predetermined by the technology itself; it may depend on whether the system is designed to substitute for, or to preserve, the user's understanding, a distinction our findings motivate and offer as a precedent for future research, though we have not tested it beyond the diabetes context.

\section{Related Work}\label{sec:related}
Our study sits at the intersection of three bodies of research encompassing five strands: ubiquitous computing, human--automation interaction, wearable diabetes technologies, long-term self-tracking, and the consequences of user disengagement. Together, they pose a single core question: \emph{when systems become increasingly capable of performing interpretive and decision-related work for users, what happens to users' own interpretive practices, autonomy, and long-term health management?} We move from the design philosophy of ubiquitous computing, through wearable diabetes computing (CGM, AID), the evolution of long-term self-tracking, and how automation reshapes cognition, to the consequences of cognitive disengagement. Each thread has developed independently; none has been linked through an eight-year longitudinal study of the same users, the empirical bridge our study provides.

Here, \emph{diabetes computing systems} denotes the configurations of digital devices, applications, and connected functions through which participants monitored and managed diabetes, including fingerstick meters, CGM and AID devices, and functions such as visualization, diet tracking, alerts, and data sharing. We reserve unqualified \emph{computing systems} for the broader class of ubiquitous, disappearance-oriented technologies of which diabetes systems are one instance, invoked in our title and in discussing generalizability beyond diabetes care. \emph{Cognitive engagement} denotes users' practices of attending to, comparing, contextualizing, interpreting, verifying, and questioning health data and system outputs; \emph{cognitive disengagement} denotes the longitudinal shift whereby continued behavioral use becomes increasingly decoupled from self-initiated sensemaking and evaluation. \emph{Stages} describe recurring configurations of system capabilities and user practices; \emph{trajectories} describe how participants move through, remain within, skip, or return to these configurations over time. The composite three-stage description, Exploration and Sensemaking, Collaboration and Delegation, and Cognitive Disengagement, we call the \emph{three-stage account}. \emph{Substitutive automation} denotes system functions performing interpretive or decision-related work users previously undertook themselves.

Because these patterns may arise when other ubiquitous systems increasingly perform work previously done by users, our analysis may sensitize examination of smart thermostats, driver-assistance systems, fitness trackers, and related technologies, though whether the same relationships hold outside diabetes management remains an empirical question rather than an assumed property of automation.

\subsection{Ubiquitous Computing, Cognitive Offloading, and Human--Automation Interaction}

Weiser's vision in \emph{The Computer for the 21st Century} framed the most profound technologies as those that ``disappear'' into the fabric of everyday life \cite{weiser1991computer}. Later researchers have criticized the calm-computing orientation as reactive \cite{rogers2006moving}, unrealistic \cite{bell2007yesterday}, and lacking agency \cite{caceres2012ubicomp}. These critiques are conceptually compelling but empirically thin, showing disappearance \emph{could} be problematic without explaining how consequences develop as practices and capabilities co-evolve. The unresolved issue in long-term health management is not acceptance or rejection but whether a system stays continuously used while users become less involved in interpreting its outputs. Our study offers longitudinal evidence for this, without assuming disappearance necessarily produces disengagement.

This surfaces a key issue: long-term offloading of cognitive work to automated systems, which occurs when people use actions or external resources to reduce a task's processing demands \cite{risko2016cognitive}. External reminders, records, and computational tools reduce memory, attention, and calculation demands, often improving immediate performance \cite{wei2026differences,grinschgl2021consequences,peng2025cognitive}; interface design shapes how much offloading occurs \cite{grinschgl2020interface,zhang2025slideaudit}, and believed reliability modulates it \cite{weis2019using}. This research has largely examined whether and when people delegate particular tasks and their performance with external support \cite{grinschgl2020interface,jin2024mobile}, with less insight into how offloading develops over years as the system itself grows more capable. We therefore know little about whether sustained offloading changes the frequency and form of users' interpretive practices, whether they stay practised and motivated enough to resume them, and how this unfolds among lay users in everyday health management.

Cognitive disengagement is not simply greater offloading. Offloading redistributes particular demands to an external resource and can remain compatible with active sensemaking, reducing repetitive memory or calculation demands without replacing interpretation of harder tasks \cite{peng2025cognitive,meng2026does}. \emph{Disengagement}, by contrast, is the longitudinal decoupling of continued use from self-initiated sensemaking: users keep wearing the device and responding to outputs, while attention becomes increasingly triggered by the system rather than self-initiated. Sustained offloading may contribute to this when it extends from discrete tasks to substituting for interpretive work, but does not inevitably cause it. It also differs from automation bias, over-reliance, and trust: bias concerns decision errors from following cues without verification, whereas disengagement need not involve an error and may stem from habituation or broader contextual factors \cite{skitka1999does,romeo2026exploring,goddard2012automation}; over-reliance is typically accepting an incorrect recommendation, whereas disengagement is a reduction in self-initiated interpretation, contextualization, verification, and questioning that can occur even with appropriate reliance and reliable performance \cite{buccinca2021trust,lu2024does}; trust concerns attitudes about a system's ability to support goals, whereas disengagement concerns how users engage with outputs in practice \cite{lee2004trust,bussone2015explanations,zhang2025selective,li2023modeling}. Trust may reduce verification and disengagement may correlate with bias or over-reliance, but is defined by neither.

Our study extends this literature into a longitudinal, lay-user, implicit-automation setting, examining how interpretive practices, modeling data, judging bodily state, evaluating output, may become less frequent as capability and use accumulate. Such reduction needs no AI-labeled system or deliberate handover; it can develop through daily repetition. We additionally introduce a multi-stakeholder design-side perspective involving clinicians, designers, and vendors that links longitudinal changes in users' cognitive engagement to the practical reasoning through which substitutive systems are produced.

\subsection{Long-Term Personal Informatics and Wearable Diabetes Technologies}
Continuous glucose monitoring (CGM) has fundamentally reshaped diabetes self-management, replacing intermittent measurements with continuous, high-frequency data, augmented by alerts and predictions \cite{lawton2018patients,rodbard2016cgm}. Pervasive-healthcare research has also argued for embedding such sensing in everyday contexts \cite{bardram2008pervasive}. 
The literature, however, does not only present a simple empowerment narrative \cite{lawton2018patients,messer2020older,bartolome2021glucomine,funnell2004empowerment}. Although continuous sensing may increase awareness of glucose patterns and support self-management, it also changes the work required of users, including how they interpret continuous data, respond to system-generated feedback, and determine when system outputs require further examination.

Recent work focusing on automated insulin delivery (AID) shows improved clinical outcomes and reduced cognitive burden, but also introduces new tensions like trust in algorithms, fear of failure, and loss of control \cite{barnard2015psychosocial, kovatchev2019closedloop, lawton2020clinician}. The patient-led DIY-AID community (e.g., OpenAPS \cite{lewis2016openaps}) and CSCW research on lived data and care work \cite{kaziunas2018lived} have further shown that closed-loop systems are entangled with collaborative reasoning, tinkering, and emotional labor that clinical evaluations alone cannot capture. These studies demonstrate that automation does not simply remove patient work but redistributes it across users, devices, family members, and clinicians. However, much of this research examines experiences surrounding adoption, implementation, or a bounded period of use rather than changes in the same users' interpretive practices across successive generations of diabetes technologies.

Personal informatics research has increasingly recognized that tracking practices and purposes change over time. Meyer et al.\ distinguish very-long-term tracking from shorter-term tracking and show that tracking conducted over years or decades may move between purposeful and incidental forms \cite{meyer2020life}. Niess and Woźniak's Tracker Goal Evolution Model similarly shows that tracking goals are not fixed but may evolve as users' needs, reflection, and trust change \cite{niess2018supporting}. Longitudinal and process-oriented studies further demonstrate that users may appropriate trackers for goals that depart from designers' assumptions, as in Homewood's 18-month autoethnography of using a fitness tracker to reduce rather than increase activity \cite{homewood2023self}, and that wearable data may become part of an extended journey involving bodily awareness, interpretation, and clinical diagnosis \cite{keys2024think}. Together, this work establishes that temporality matters, but it primarily characterizes changing tracking purposes, goals, or episodes of appropriation rather than how increasingly substitutive system functions reshape users' self-initiated interpretation over multiple technology generations.

In addition, personal informatics research offers an essential conceptual vocabulary for analyzing our findings, because CGM is also a type of personal informatics tool. Li, Dey, and Forlizzi's stage-based model decomposes self-tracking into preparation, collection, integration, reflection, and action, identifying \emph{reflection} as the stage at which understanding is forged \cite{li2010stagebased}. The model surfaces a fundamental distinction: \emph{having} data is not the same as \emph{understanding} it; understanding requires users to invest active cognitive work.Similar patterns appear in algorithmically curated data summaries, where compression can limit users' own reflection \cite{li2026platform,meng2026data}. Rooksby et al.\ reframed tracking as \emph{lived informatics}, emphasizing that practice is non-linear, socially embedded, and frequently interrupted \cite{rooksby2014lived}. Epstein et al.\ extended this with a lived-informatics model that treats tracking as a dynamic, situated practice rather than a linear stage sequence \cite{epstein2015lived}. Choe et al.'s study of Quantified Self practitioners highlighted recurring pitfalls, including tracking too many variables and losing situational judgment \cite{choe2014understanding}. Their work on semi-automated tracking further argues that manual and automated data collection should be combined according to the purpose of tracking, the feasibility of data capture, and users' motivation, so that automation can reduce capture burden without removing awareness and participation \cite{choe2017semi}. This perspective is particularly relevant to functions such as food logging and data sharing, which may require users to provide contextual information or communicate with others even when glucose sensing and calculation are increasingly automated. Such functions therefore represent potential sites of continued cognitive engagement rather than equivalent instances of full delegation.

Most relevant to our work is research on what happens \emph{after} active tracking. Epstein et al.\ analyzed lapsing and abandonment of personal-informatics tools and showed that ``stopping'' is rarely a clean exit but often a partial disengagement \cite{epstein2016beyond,jones2021determinants}. Clawson et al.\ used ``abandoned'' trackers as a window onto the misalignment between evolving user needs and system designs \cite{clawson2015noaccount}. Kabir and Wiese's meta-synthesis further organizes barriers and facilitators reported across personal informatics research, demonstrating that engagement is shaped by interacting user, system, data, and contextual conditions rather than usability alone \cite{kabir2023meta}. Adjacent strands document the invisible \emph{data work} of patients with chronic conditions \cite{ancker2015reminded} and have produced reusable design patterns for habit support \cite{stawarz2015beyond,consolvo2008activity, doyle2014older} along with methodological frames for evaluating health-behavior-change technologies \cite{klasnja2011how,meng2026tibetcpr,meng2026engagement}. In the diabetes domain, Mamykina and colleagues built MAHI and argued for a \emph{sensemaking} perspective on chronic-disease self-management, foregrounding the cognitive work users must do to turn data into actionable understanding\cite{mamykina2008mahi, mamykina2015sensemaking, mamykina2017datadriven}—building on the older PARC sensemaking tradition that frames information work as cost-structured and representation-driven \cite{russell1993sensemaking,pirolli2005sensemaking} In particular, Mamykina et al.'s account of personal discovery shows how people use self-monitoring data to identify associations between everyday behaviors and glucose changes, providing a close conceptual precedent for the exploration and sensemaking practices examined in our first stage \cite{mamykina2017personaldiscovery}.

These threads provide an indispensable foundation, yet leave a critical conceptual gap. Existing abandonment research presumes a continuum from active use to behavioral exit, with attention focused on \emph{when} and \emph{why} users stop. Lived-informatics frameworks acknowledge non-linearity but typically span weeks to roughly two years, too short to capture structural change at longer time scales. Sensemaking frameworks center cognitive work but do not track how users' participation in that work may diminish as system functionality grows. Our longitudinal data reveal a pattern these frameworks cannot fully describe: \emph{continued use with cognitive disengagement}---the device is still worn, the data pipeline still runs, the user has not behaviorally abandoned the system, yet active interpretation, model-building, and questioning have become markedly less frequent. This is neither active engagement nor abandonment but a third state---behavioral retention combined with cognitive disengagement---that emerges through gradual sedimentation rather than a single moment of change. Identifying and theorizing this state is our central contribution to the personal-informatics conversation.

\subsection{Potential Risks of Cognitive Disengagement in Wearable Health Systems}
Automation and cognitive offloading reduce the immediate demands of chronic-disease management, but may coexist with risks once continuous sensing and automated interpretation become the primary basis for monitoring and action: failures of health and safety judgment, information and alert fatigue, reduced situation awareness and recovery capacity, diminished agency, emotional burden, redistributed care work, weakened attention to data practices, and unequal vulnerability across users. We treat these as plausible, interacting consequences rather than inevitable outcomes.

\textit{Health and safety risks.} Systems may monitor routine cases successfully while leaving users responsible for rare, abnormal situations. Bainbridge's "irony of automation" holds that users must intervene precisely when automation fails, despite fewer opportunities to practise the relevant judgment \cite{bainbridge1983ironies}. The out-of-the-loop performance problem shows automation shifts users to passive processing, reduces situation awareness, and impairs resumption of manual control after failure \cite{endsley1995out}. Automation misuse and bias describe users following outputs beyond their reliability or failing to seek contradictory information \cite{parasuraman1997humans,skitka1999does,goddard2012automation,lu2021human}. Explanations alone may not prevent overreliance; forcing analytic engagement reduces acceptance of incorrect advice, at the cost of effort \cite{buccinca2021trust}.

This matters for wearable diabetes systems, where sensing errors, signal loss, incorrect settings, communication failures, or unperceived alerts directly affect health action. CGM/AID studies report concerns over inaccurate or delayed readings, signal interruption, system failure, algorithmic limits, and uncertainty about delegating control \cite{lawton2018patients,barnard2015psychosocial,lawton2020clinician}, but mainly during particular periods of use, leaving unclear whether years of routine success erode readiness to notice abnormalities, question outputs, or act when protection fails.

\textit{Information fatigue, alert fatigue, and attentional withdrawal.} Continuous monitoring can produce more data and prompts than users can process. Personal-informatics research documents difficulties from excessive variables, ambiguous patterns, and reflection fatigue \cite{choe2014understanding,epstein2016beyond,meng2026ubiquitous}. Kabir and Wiese show engagement barriers arise from interacting user, system, data, and context factors rather than usability alone \cite{kabir2023meta}, so disengagement can reflect low perceived informational value rather than low motivation.

Notification research shows recurring prompts compete for attention: users receive substantial notification volumes linked to stress and overwhelm, and routinely defer, dismiss, or categorize prompts by importance \cite{pielot2014situ,sahami2014large,pielot2017beyond}; rising volume adds management work, and users value deferral and prioritization tools \cite{telang2025can}. Even lightweight smartwatch prompts see non-response varying by time, activity, mobility, and context \cite{ponnada2022contextual}. In safety-critical systems, repetitive or poorly prioritized prompts risk habituation and "notification blindness," obscuring consequential signals \cite{shivers2013turn}.

\textit{Reduced situation awareness, interpretive readiness, and recovery capacity.} Beyond fatigue, fewer opportunities to practise interpretation matter: passive monitoring supports situation awareness more weakly than active processing and can delay intervention when automation fails \cite{endsley1995out,parasuraman2010complacency}. Reflection and sensemaking connect data to circumstances and build usable understanding \cite{li2010stagebased,mamykina2015sensemaking}; semi-automated tracking research argues automation should cut unnecessary capture work without removing participation that sustains awareness \cite{choe2017semi}. The risk is less a discrete lost skill than reduced interpretive readiness, users comparing, contextualizing, verifying, and questioning less often, and resuming these more slowly when systems behave unexpectedly.

\textit{Privacy, data agency, and unequal vulnerability.} Continuous sensing raises risks around data visibility, sharing, secondary use, and control. Human-data interaction research calls for systems that remain legible, negotiable, and open to intervention \cite{mortier2014humandata,meng2026living}, with contextual-integrity and mobile-health privacy frameworks as complementary lenses \cite{muchagata2018wearable,nissenbaum2004contextual,tovino2020wearable}. Privacy expectations vary by data sensitivity, collection context, recipient, and purpose, while users often lack clear information about these flows \cite{motti2015users,naeini2017privacyexpectations,caine2013granular,meng2026neglect}; reduced engagement may make such flows less salient.

These risks are usually studied separately. Existing work rarely examines how they connect over years of use as systems shift from displaying data to interpreting, predicting, and acting for users, or how users continue wearing and responding to a system while progressively reducing self-initiated interpretation. Our longitudinal study addresses this gap.

\section{Method}\label{sec:method}

\subsection{Study Design}
This study adopts a longitudinal qualitative research design, aiming to understand users' behavioral changes and the evolution of their interpretations during long-term use of diabetes computing systems. The study is conducted around the same group of primary participants (P1-P37), with multiple rounds of interviews carried out at five time points-2017, 2019, 2021, 2023, and 2025-in order to capture the continuity and transformation of their usage contexts, interpretive approaches, and decision-making logics across different stages. This design enables the study to go beyond observations at a single time point and to identify the trajectories of change that individuals experience during prolonged interaction with the system.

On this basis, the study further introduces doctors, designers, and vendors as supplementary participants, and collects explanatory information from the system design and deployment side through one-time interviews. The two types of data correspond to each other in the analysis: the former is used to reveal dynamic changes on the user side, while the latter is used to explain how the system is constructed and how its usage is preconfigured, thereby forming an overall understanding of the relationship between users and diabetes computing systems. Full details of the design rationale, including the choice of wave intervals, dual-group logic, sampling strategy, and wave-by-wave technology context, are provided in Appendix~\ref{app:studydesign}.

\subsection{Participants}
This study includes two participant groups. The primary group (P1--P37) consists of 37 individuals (21 male, 16 female), aged 39--70, from both rural (n=19) and urban (n=18) settings. Their educational backgrounds show clear diversity, including semi-literate, primary school, junior high school, high school, and higher education (bachelor's, master's, and doctoral degrees). Their occupational distribution is also diverse, covering service workers, skilled laborers, farmers, retail workers, caregivers, as well as unemployed individuals. Such differences in education level, occupational structure, and living environment enable this study to capture the diverse experiences of different populations in encountering and responding to diabetes computing systems. See Appendix~A for details.

In addition, 15 practitioners provide supplementary perspectives: 5 doctors (Doctor1--Doctor5), 5 designers (Designer1--Designer5), and 5 vendors (Vendor1--Vendor5). The doctor participants all hold doctoral degrees in medicine with rich clinical experience; the designers hold at least a bachelor's degree with backgrounds in product or system design; the vendors cover different levels of education and business experience. This group supplements understanding of system design logic and deployment considerations, forming a cross-perspective analysis. See Appendix~B for details.

Primary participants were recruited through in-situ contact and snowball sampling, emphasizing richness and diversity of life experiences. Practitioner participants were recruited through purposive sampling, with balanced numbers across roles to avoid any single perspective dominating.

\subsection{Ethics}

This study received institutional ethics approval. All participants provided informed consent, could withdraw at any time, and received 50 RMB compensation per interview. Data were anonymized during recording and analysis, stored securely, and used solely for academic research. Interviews used open, non-leading questioning and were paused when necessary to ensure participants' comfort. Full details of data management and security procedures are provided in Appendix~\ref{app:datacollection}.

\subsection{Data Collection}

This study conducted semi-structured interviews with primary participants at five time points (2017, 2019, 2021, 2023, 2025), each lasting 50--73 minutes, one-on-one, audio-recorded and transcribed with informed consent. Researchers avoided evaluative or leading language and focused on probing concrete situations and behaviors. Full details of interview logistics, longitudinal consistency protocols, interviewer conduct, participant retention strategy, and data management are provided in Appendix~\ref{app:datacollection}.

For patient interviews, the same protocol was used across all five waves. Each interview moved through four phases: (1)~the context of technology entry into the participant's life; (2)~a recent concrete usage episode, probed for trigger, context, behavioral response, and delay/neglect; (3)~daily usage patterns, including proactive checking, responses in the absence of alerts, and perceived information burden; and (4)~long-term behavioral change, data-flow awareness, and open-ended reflection on system expectations. The full interview guide is provided in Appendix~C.

Interviews with doctors, designers, and vendors were conducted only in 2025, following a parallel semi-structured approach focused on design and decision-making practices: observations of actual user behavior, specific design trade-offs (risk control, user burden, technical constraints), and the practical logics generating current system forms. The full guide is provided in Appendix~D.

\subsection{Data Analysis}
For primary participants (P1--P37), we used longitudinal trajectory analysis \cite{saldana2003longitudinal,grossoehme2014longitudinal,thomson2003making, tuthill2020longitudinal}, conducted jointly by three researchers (R1, R2, R3) over 14 months (Jan.~2025--Feb.~2026). The basic coding unit was the \emph{participant--wave--episode}: a contiguous narrative segment describing one concrete usage episode in one interview wave. A total of 1,182 episodes were coded across the five waves using an iteratively refined codebook (47 codes, six families). Three-way Fleiss' $\kappa$ on a 20\% random sub-sample (8 case files, 246 episodes) reached 0.81 after dispute resolution; details of the codebook development, inter-rater procedure, dispute-resolution protocol, and saturation assessment are provided in Appendix~G.

To distinguish stable trajectory shifts from situational fluctuation, a coded change had to satisfy four criteria: persistence across at least two consecutive waves, retrospective re-narration by the participant, a corresponding behavioral correlate in usage episodes, and robustness across contexts (work, sleep, exercise, illness). After per-participant trajectories were finalized, cross-case synthesis \cite{ayres2003within} identified recurring patterns, with deliberate attention to disconfirming and partial cases (\S4.4).

For supplementary participants (doctors, designers, vendors), we used interpretive thematic analysis \cite{braun2006using,nowell2017thematic} with the same three coders and dispute-resolution protocol (Cohen's $\kappa$ on a 30\% sub-sample: 0.69--0.78). The resulting interpretive structures are not presented as independent results but are used to reveal alignment and mismatch between design logics and user practices.

\section{Results}
\subsection{The Evolutionary Process of Technology}

Overall, during the research period from 2017 to 2025, users' technology use went through three broad periods of evolution. (1) 2017--2019: CGM gradually entered users' lives, and physiological data began to be displayed on-screen; the earlier model of fingerstick blood glucose testing combined with apps that provided only general knowledge gradually faded away. (2) 2019--2021: Real-time feedback designs based on CGM gradually matured, and part of the information-processing work shifted from reliance on physicians and patient self-querying toward system-mediated delegation. (3) 2021--2025: CGM-based data began to be used by systems for prediction, recommendation, and even automated decision-making (AID). Overall, each period builds on the accumulation of the previous period, with the system progressively taking over cognitive work that had previously been performed by the user.

It is worth noting that, although most participants' technology-adoption timelines broadly aligned with these three periods, the three periods described above are only a macro-level distinction; there are no clear boundaries between them, and the specific timing of technology adoption varies considerably from individual to individual. A small number of participants deviated from these three time windows; for example, some urban residents and highly educated participants (e.g., P22, P23), as well as patients with particularly severe disease (e.g., P32), tended to adopt CGM technology and gain access to real-time feedback apps earlier, while participants from rural areas, with lower education levels and worse economic conditions (e.g., P27) tended to adopt these technologies later. In addition, a small number of participants, because they began using the technology relatively late, did not experience long periods in first and second periods. For example, when P29 began using CGM in 2024, the technology was already fairly mature, and much of the content of first and second periods had already been integrated into the glucose-monitoring system. Detailed technology-use information is shown in Table 1.

\begin{table*}
  \caption{Technology Use Over Time}
  \label{tab:tech_use}
  \centering

  \begingroup
  \scriptsize
  \renewcommand{\arraystretch}{1.0}
  \setlength{\tabcolsep}{2pt}

  \begin{tabular}{
    p{0.045\textwidth}
    p{0.175\textwidth}
    p{0.175\textwidth}
    p{0.175\textwidth}
    p{0.175\textwidth}
    p{0.175\textwidth}
  }
    \toprule
    \textbf{ID} &
    \textbf{Mid-2017} &
    \textbf{Mid-2019} &
    \textbf{Mid-2021} &
    \textbf{Mid-2023} &
    \textbf{Mid-2025} \\
    \midrule

    P1  & Fingerstick+App & Fingerstick+App & CGM (H2 2020)+App & CGM+App & CGM+App \\
    P2  & Fingerstick+App & Fingerstick+App & Fingerstick+App & Fingerstick+App & CGM (H1 2025)+App \\
    P3  & Fingerstick+App & CGM (2018)+App & CGM+App & CGM+App & CGM+App \\
    P4  & Fingerstick+App & Fingerstick+App & CGM (2020)+App+Diet Tracking & CGM+App+Diet Tracking & --- \\
    P5  & Fingerstick+App & CGM (H2 2018)+App & CGM+App & CGM+App & CGM+App \\
    P6  & Fingerstick+App & CGM (2019)+App & CGM+App & CGM+App+Data Sharing & CGM+App \\
    P7  & Fingerstick+App & Fingerstick+App & Fingerstick+App & --- & --- \\
    P8  & Fingerstick+App & Fingerstick+App & CGM (H2 2020)+App & CGM+App & AID (2024)+App \\
    P9  & Fingerstick+App & Fingerstick+App & CGM (2020)+App & CGM+App & AID (2024)+App \\
    P10 & Fingerstick+App & Fingerstick+App & Fingerstick & CGM (H1 2023)+App & CGM+App \\
    P11 & Fingerstick+App & CGM (H1 2018)+App & CGM+App & CGM+App & CGM+App \\
    P12 & Fingerstick+App & CGM (H2 2017)+App & CGM+App & CGM+App & CGM+App \\
    P13 & Fingerstick+App & CGM (2018)+App & CGM+App & CGM+App & CGM+App \\
    P14 & Fingerstick+App & CGM (2018)+App & CGM+App & CGM+App & --- \\
    P15 & Fingerstick+App & CGM (H1 2018)+App & CGM+App+Data Sharing & CGM+App+Data Sharing & CGM+App+Data Sharing \\
    P16 & Fingerstick+App & CGM (H1 2019)+App & CGM+App & CGM+App & AID (2025)+App \\
    P17 & Fingerstick+App & CGM (H1 2019)+App & CGM+App+Data Sharing & CGM+App+Data Sharing & CGM+App+Data Sharing \\
    P18 & Fingerstick+App & Fingerstick+App & Fingerstick+App & Fingerstick+App & CGM (2024)+App \\
    P19 & Fingerstick+App & CGM (H2 2017)+App & CGM+App & CGM+App & CGM+App \\
    P20 & Fingerstick+App & CGM (before H1 2019)+App & CGM+App & CGM+App & CGM+App \\
    P21 & Fingerstick+App & Fingerstick+App & CGM+App & CGM+App & AID (H1 2025)+App \\
    P22 & CGM (H2 2016)+App & CGM+App & CGM+App & CGM+App & CGM+App \\
    P23 & CGM (H2 2016)+App & CGM+App & CGM+App & CGM+App & CGM+App \\
    P24 & Fingerstick+App & CGM (H2 2018)+App & CGM+App & CGM+App & CGM+App \\
    P25 & Fingerstick+App & CGM (H2 2017)+App & CGM+App+Diet Tracking & CGM+App & CGM+App \\
    P26 & Fingerstick+App & CGM (2018)+App & CGM+App+Diet Tracking & CGM+App+Diet Tracking & CGM+App \\
    P27 & Fingerstick+App & Fingerstick+App & Fingerstick+App & Fingerstick+App & CGM (2024)+App \\
    P28 & Fingerstick+App & CGM (2019)+App & --- & --- & --- \\
    P29 & Fingerstick+App & Fingerstick+App & Fingerstick+App & Fingerstick+App & CGM (2024)+App \\
    P30 & Fingerstick+App & --- & --- & --- & --- \\
    P31 & Fingerstick+App & --- & --- & --- & --- \\
    P32 & CGM (H1 2017)+App & CGM+App & CGM+App & CGM+App & AID (2025)+App \\
    P33 & Fingerstick+App & --- & CGM (2020)+App & CGM+App & CGM+App \\
    P34 & Fingerstick+App & Fingerstick+App & CGM (2019)+App & CGM+App & CGM+App \\
    P35 & Fingerstick+App & CGM (2018)+App & --- & --- & --- \\
    P36 & Fingerstick+App & CGM (2018)+App & CGM+App & AID (2023)+App & AID+App \\
    P37 & Fingerstick+App & CGM (H1 2019)+App & CGM+App & CGM+App & CGM+App \\

    \bottomrule
  \end{tabular}

  \endgroup
\end{table*}

\subsubsection{Visualization and Continuous Sensing}

The first key technological transition was the shift from intermittent measurement to continuous monitoring, occurring mainly between 2017 and 2019. Before using CGM, participants mostly relied on fingerstick glucose meters to obtain blood glucose values; as CGM gradually entered daily use, glucose data began to be presented as continuous readings. In the Wave 2 (2019) interviews, the majority of participants (22/34) reported having experience using CGM. For example, the statements below from P12 at the 2017 and 2019 time points illustrate this process.

P12 (2017): \textit{``I use a probe myself, and the glucometer the doctor told me to buy to record my blood sugar. The glucometer is okay, it's not too big, but it still hurts a bit.''}
P12 (2019): \textit{``Before, I had to find time every day to prick my finger. Sometimes it was inconvenient outside. Now it's attached to my arm, and it's measuring wherever I go. Much easier to use.''}

At the same time, glucose-monitoring systems became increasingly integrated with smartphones, allowing users to view glucose readings and their trends over time, rather than being limited to a single measurement result. The statement from P19 below, about being able to see a changing line, and the statement from P14, about sharing data with a doctor, both illustrate this application.

P19 (2019): \textit{``Before, after measuring, it was just one number. Now I can open my phone and see a whole line, changing from morning to night. It's very convenient, I can see abnormalities at a glance.''}
P14 (2019): \textit{``Before going to the hospital, I can bring out the data from the past few days for the doctor to see, not just one number from that day.''}

In addition, 12 users had still not adopted CGM by Wave 2 (2019). The main reasons included clinical advice (P7: ``Mild diabetes doesn't require invasive monitoring''), concerns about wearing the device continuously (P9: \textit{``I was worried about the possible health effects of continuous skin puncture''}), economic reasons (P27: \textit{``It's still somewhat hard for me to afford CGM''}), and insufficient awareness of CGM among the participant themselves or their community (P29: \textit{``My doctor seems to have mentioned CGM before, but our community is fairly underdeveloped, and my doctor isn't very familiar with it either.''}). In the course of the subsequent study, aside from P7, who did not need CGM, the other participants who had not yet adopted CGM due to economic factors, digital literacy, or the urban-rural gap also came into contact with CGM technology at various later points. P9 gradually stopped worrying about this issue in later interviews.

Finally, it is worth noting that a small number of participants with higher educational backgrounds adopted CGM earlier than most participants, for example P22: \textit{``Oh, you mean this: I actually started using CGM around the end of 2016, though the technology was still fairly primitive back then.''}

\subsubsection{Real-Time Feedback and Behavior Triggering}
The second key technological transition was the rise of real-time feedback functions built on top of CGM, occurring mainly between 2019 and 2021. The difference from the previous  period is that users' phones and other devices began to automatically process and interpret the data collected by CGM, providing different feedback depending on the situation. The main difference from the next  period is that this period still centered mainly on real-time glucose tracking and direct feedback, rather than in-depth analysis, prediction, or automated decision-making.

For example, P1 below describes how, when his blood glucose dropped too low, the phone app alerted him that something was wrong, prompting him to take action.

P1 (2021): \textit{``Once on the subway, my phone kept vibrating. I thought someone was contacting me, but it turned out my blood sugar was dropping, so I had to get off and buy something to eat.''}

In addition, remote data-sharing emerged during this period: glucose data could be synchronized to the cloud so that family members or doctors could view it and provide feedback. Although the feedback was not necessarily generated directly by the technological system itself, the sharing technology provided a platform for real-time feedback. The statements below from P15 and P17 describe how doctors and family members provided feedback upon observing abnormalities.

P15 (2021): \textit{``Once I didn't notice my low blood sugar myself, but my sister called me first asking if I was feeling unwell. She had already seen the data dropping on her side.''}
P17 (2021): \textit{``Now before I go for a follow-up visit, the doctor has already seen my recent data. When I walk in, they immediately start pointing out the problems.''}

In addition, toward the end of this period, a small number of participants mentioned that the functions of mobile apps gradually expanded to include statistical analysis and dietary and exercise recommendations. By having users upload photos or text information, the app could track dietary intake and, combined with CGM data, provide recommendations. The statements below from P25 and P26 describe the app functions of this period.

P25 (2021): \textit{``Actually, the app's functions are quite limited: a lot of foods don't have sugar-content data available. But I find it interesting that after eating, you can see the blood glucose data in the app: the app will show you the food you ate along with reminders about abnormal glucose readings.''}

P26 (2021): \textit{``Now I basically open the app after meals to take a look. Sometimes I also record what I ate, and the app calculates how much sugar is in the food.''}

Finally, it is worth noting that in the Wave 3 (2021) interviews, P9, P34, and others also mentioned their experience of starting to use CGM (later than the overall pace of the participant group), while a small number of participants, such as P23, had already begun discussing content belonging to the next period, app-based prediction, (``The app gives me a probability prediction based on my past data'').

\subsubsection{Prediction, Recommendation, and Automated Decision-Making}
The third major technological transition was the point at which systems began to predict potential glycemic events based on real-time data streams, rates of change, and historical patterns, occurring mainly between 2021 and 2025. For example, P8 below describes a scenario in which the system predicted he might be about to experience hypoglycemia, while P16 describes adjusting his own behavior based on predicted trends in blood glucose.

P8 (2023): \textit{``Once it warned me in advance that I might go low and suggested I eat a little more, but I was busy and didn't notice. Later my blood sugar dropped quickly and I felt very dizzy.''}
P16 (2023): \textit{``Now sometimes it's not about waiting for an alarm. If I see it predicting a rise, I eat a bit less in advance. Sometimes I tell the app I'm going to play badminton, and it suggests I eat a bit beforehand.''}

In addition, automated insulin delivery (AID) pushed the system's functions further toward automated decision-making: AID integrates CGM, an insulin pump, and a control algorithm into a partially autonomous regulatory loop. For participants using this type of system, glucose monitoring, medication delivery, and algorithmic decision-making were integrated into a single continuous operating loop. For example, P32 mentions how AID reduced his own cognitive burden.

P32 (2025): \textit{``Before, I had to calculate everything before each meal. Now it automatically adjusts part of it, and I just need to confirm roughly.''}
Multi-device integration also became widespread, with glucose data combining with exercise, sleep, and heart rate data across platforms:

It is also worth noting that, by the end of the study in 2025, all 29 participants who had fully completed all five waves had CGM experience, while a total of 6 participants were using AID.

\subsection{User Usage Evolution}
Users' modes of use and cognitive patterns also displayed three corresponding stages of evolution, broadly matching the three periods of technological evolution described in Section 4.1. Early on, users actively explored the data and tried to understand changes in their bodies (corresponding to the emergence of continuous monitoring and visualization technology described in 4.1.1); subsequently, an increasing amount of data-interpretation work came to be taken over by the system, and users began to rely more on judgments provided by the system (corresponding to the gradual maturation of information processing and real-time feedback described in 4.1.2); after long-term use, active interpretation and sustained attention further diminished, gradually forming an interaction pattern centered on system monitoring and user response, that is, cognitive disengagement (corresponding to the further development of the glucose-monitoring system with prediction and automated processing functions, as described in 4.1.3).

\subsubsection{Stage 1: Exploration and Sensemaking}
In the early stage during which CGM entered users' lives (corresponding mainly to the first period of technological evolution described in 4.1.1, 2017--2019), participants mostly displayed an ``understanding-oriented mode of use,'' for example, frequently checking their data, comparing changes across different situations, and linking fluctuations to specific behaviors. During this stage, users tended to view the glucose-monitoring system as a medium through which to observe their own bodies.

P17 (2019): \textit{``At that time, I deliberately made comparisons, like eating similar things today and yesterday, and seeing whether the changes were the same.''}

P17's behavior represents users' exploration of the glucose-monitoring system itself during this period, and most participants also mentioned cross-validating the data, even actively verifying it when the output conflicted with their own bodily sensations. For example, P24 mentions using the older fingerstick method to verify the glucose values given by CGM.

P24 (2019): \textit{ ``Sometimes the value doesn't match how I feel, so I use the old finger-prick method to check again.''}

This process of cross-validation demonstrates that participants actively engaged in independent interpretation, and the trajectories that follow document how this behavior became gradually less frequent over time.

In addition, as mentioned in Section 4.1, while a small number of participants had not yet adopted CGM during the early period of the study (2017--2019), most users, once they later adopted CGM, tended to display the same kind of curiosity, understanding, and sense-making, for example, P27 (\textit{``How is the CGM data actually generated?'' [2025]}) and P18, who mentioned that he would \textit{``try to infer his own physical condition based on the data.''} However, because these participants encountered CGM-based feedback and decision-making technologies at a later point, when those technologies were already more mature, they tended to show less verification and less questioning.

It is worth noting that a small number of participants did not fit the overall direction of this stage. Some participants, during this stage, tended  to treat the glucose-monitoring system as an authority to be trusted directly, without displaying much evident curiosity, understanding, or reflection.

P11 (2019): \textit{``It's just so convenient; I don't have to think about anything anymore, it (the CGM system) does it all for me. It's really great. Now I can just show the correct numbers to the doctor, and I don't have to bother recording and checking things every single day.''}

\subsubsection{Stage 2: Collaboration and Delegation}
As usage time increased, and as the automated real-time feedback and alert technologies described in 4.1.2 gradually matured between 2019 and 2021, most users shifted from ``actively interpreting the data'' to ``directly responding to the system's conclusions.'' This shift was typically gradual: users did not consciously decide to stop interpreting the data; rather, as the system's output became increasingly directive, the effort of independent reasoning no longer seemed necessary or effective.

P19 (2021): \textit{``Later, it's no longer like the beginning where I think about why. It directly tells you, like whether it's rising or getting high, and then I know to pay attention.''}
P14 (2021): \textit{``At the beginning, you think a lot about why, but later you feel that it has already judged for you, and you just need to follow it.''}

P19's statement reflects a reduction in reflective thinking, and P14's statement reflects compliance with the glucose-monitoring system; both indicate that users' active thinking was declining, and that interpretive authority was being partially ceded.

This ceding of authority was more often experienced as a form of ``relief'' rather than as a loss: participants often described this shift in positive terms (for example, P14's use of words like ``convenient'' and ``no need to worry''). During this stage, most participants did not question this transfer of authority, and even a small number of highly educated, technologically literate participants were rarely an exception.

However, it is worth noting that cognitive ceding during this period was still incomplete. For example, the data-sharing technologies used by P15 and P17, and the food-tracking technology used by P25 and P26, as mentioned in Section 4.1.2, suggest that new, active forms of interaction, thinking, and action still emerged between 2019 and 2021. This reflects the fact that most users at this stage had not yet fully ceded their cognitive engagement. The statements from P26 and P17 below both reflect active behavior during this period, which can also be seen as a continuation of the active behavior of users from the previous stage described in 4.2.1.

P26 (2021): \textit{``The way it works is actually really simple: you just do what the app tells you: take a photo, and upload it to the app, and the app calculates it for you.''}

P17 (2021): \textit{``Having the system help me ask the doctor directly, and having the doctor just answer; I think that's a really great design.''}

\subsubsection{Stage 3: Cognitive Disengagement}
After long-term use, even though the device continued to be worn, most users' active attention to and interpretation of the data gradually declined (corresponding broadly to the final period of technological evolution described in 4.1.3, 2021--2025). Compared with the early stage, when users frequently checked their glucose curves and analyzed the reasons behind changes, most users came to rely increasingly on continuous monitoring by the system, taking action only when they received an alert or prompt. The device continued to participate in daily life, but users' own investment in interpreting the data clearly decreased.

P3 (2025): \textit{``Now I basically don't check it anymore. Unless it alarms, I won't open the interface. I just let it run in the background.''}
P6 (2025): \textit{``It keeps monitoring there, and I feel it will watch for me. If there's a problem, it will tell me. I don't look at the specific numbers like before.''}

P3 emphasizes that she almost only opens the interface when the system alarms; P6 states that continuous monitoring has replaced his habit of frequently checking specific values. Together, these changes show that users are increasingly leaving continuous monitoring and anomaly detection to the system, intervening themselves only after the system issues a signal.

One important mechanism accelerating this phenomenon is information fatigue: the continuous stream of data becomes difficult for users to process, leading them to actively reduce their engagement. P17 below raises the issue of information overload, P24 mentions feeling worn out, and P26 mentions frustration and compromise arising from being unable to fully meet every goal.

P17 (2023): \textit{``Sometimes there's too much information, always changing. Even if you look at it, you don't know how to judge. So later I just stopped checking.''}
P24 (2025): \textit{``At the beginning, I looked at it very seriously. Later it felt tiring, having to think about these changes every day, so I didn't want to deal with it.''}
P26 (2025): \textit{``Sometimes it reminds me, but I don't deal with it immediately. It feels annoying because there's too much data to evaluate, and sometimes they contradict each other. No matter what I do, there are still errors, so I just let it be.''}

However, a different set of voices emerged among a small number of participants at this stage. Although they described facing considerable challenges, certain factors triggered them into action. For example, P9 mentions the impact of an AID-related dangerous incident, while P23 mentions unease about ``handing everything over to the machine.''

P9 (2025): \textit{``After that incident (the participant received an insulin overdose due to an AID malfunction), I found I had done a lot of thinking. Machines can make mistakes, and if you hand everything over to a machine, once something this important goes wrong, it threatens your life. Now, when I feel unwell, I check my glucose data myself, even though the doctor told me this particular AID issue probably won't happen again.''}

P23 (2025): \textit{``I've also been reflecting lately. I think there's nothing wrong with taking the machine's advice, but we need to have our own judgment, our own understanding, at the very least, a spirit of skepticism. Otherwise, if you believe everything unquestioningly, it could lead to serious health problems.''}

In addition, P22's statement reflects considerable depth of thought, raising questions about the relationship between users' cognitive capacity, cognitive load, and machines, and expressing opposition to fully ceding cognition to the system.

P22 (2025): \textit{``I think that over the past few years, the integration of glucose data with the interface, and the recommendations given by the interface design, have all been fine. But I've had some reflections recently: if machines make us understand nothing at all anymore, to the point where we can't even recognize the letters `T1DM,' then has our trouble actually decreased or increased? I think that just to make sure we can still recognize `T1DM' down the road, we can't hand absolutely everything over to the machine.''}

\subsection{Structural Emergence of Latent Risks}

As users delegate interpretive authority to the system, two important types of latent risks emerge.

\subsubsection{Health Risk}
With long-term use, some users have come to rely increasingly on the judgments provided by the system, while less often actively drawing on their own experience to interpret changes in blood glucose. Compared with the early stage, when they focused on specific values and the reasons behind changes, they now more often decide whether to take action based on system prompts. Some participants also stated that, when they do not receive an alert, they typically assume by default that their current state is normal.

P20 (2025): \textit{``I don't judge whether the machine's value is reasonable. What it measures should be correct. How could I surpass a machine in understanding?''}

While the system is functioning normally, this reliance can reduce the burden of continuously interpreting the data placed on the user, but when the system fails to provide effective prompts, it can also introduce new risks.

P15 (2025): \textit{``If it doesn't remind me, I usually don't pay special attention. I assume it's normal.''}

Researchers found that P15's device volume was set to minimum while P15 had severe hearing impairment---many alerts were issued but never perceived. P15 treats ``not having received an alert'' as an important basis for judging her current state, but her experience shows that when alerts fail to be effectively communicated, this mode of judgment can lead to delays in detecting abnormal conditions.

As described in Section 4.2.3, only a small number of highly digitally literate participants (P22 and P23), along with P9 (who experienced a safety incident resulting from cognitive ceding), held clearly critical views regarding the idea that ``handing decision-making entirely over to the machine may create health problems''; most users did not perceive any health risk in this regard.

\subsubsection{Information Load}

Although technological systems are generally designed to reduce the burden of continuously interpreting data placed on users, over the course of long-term use, some users still need to continuously receive and process the prompts and information generated by the system. Sustained attention and judgment introduced a new kind of burden.

P17 (2025): \textit{``It's always changing. Sometimes it's normal, then later it changes again. It gets annoying, you don't know whether you should keep checking it.''}
P16 (2025): \textit{``At the beginning, I carefully looked at every change. Later it felt tiring because I had to check many times a day and think about whether to adjust something each time.''}

P17 believes that frequently changing data makes it hard to judge whether continuous attention is needed; P16 mentions a process of reducing investment due to the fatigue brought on by sustained attention. This burden is fragmented and continuous, making it difficult to notice and manage. Some users have begun reducing their cognitive investment by checking less frequently or ignoring certain prompts.

P26 (2025): \textit{``There's too much to do, it's impossible to do everything right. So I just let it be. As long as it's not the most serious alerts, the rest doesn't matter.''}

This coping strategy further reinforces disengagement: when the cost of processing becomes too high,  users reduce their engagement rather than resuming their earlier practice of interpretation. Even as the system lowers the barrier to understanding, it locks users into a continuous cycle of response; users ``understand less'' while ``responding more.''

\subsection{Variation, Subgroup Patterns, and 
Disconfirming Cases}

Classification and variation. Of the 37 participants, 29 completed all five interview waves (All 29 participants had experience using CGM technology). Of these, 19 (65.5\%)  showed trajectories that broadly moved through the three-stage account proposed in Section 4.2: Exploration and Sensemaking, Collaboration and Delegation, and Cognitive Disengagement, with P3, P6, P17, P19, P20, P24, P25, and P26 as the clearest examples. Information regarding the stages, classifications, and representative users is shown in Figure 1.

\begin{figure}[htbp]
    \centering
    \includegraphics[width=0.8\textwidth]{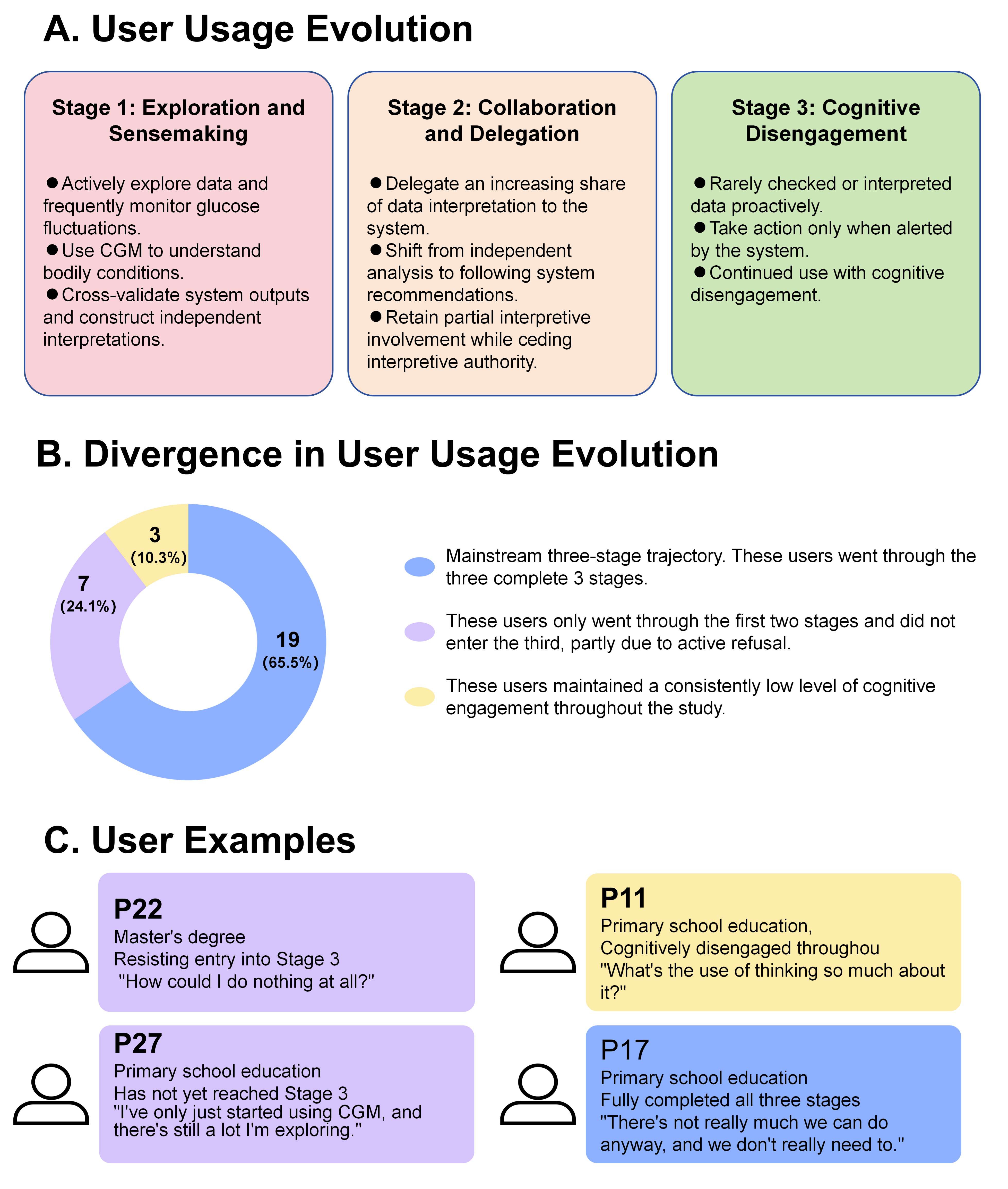}
    \caption{A: A brief overview of the three stages of usage evolution exhibited by the majority of users; B: The differentiation results for the 29 participants who completed all five interviews; C: Representative examples of specific users}
    \label{fig:figure1}
\end{figure}

Seven participants (24.1\%) showed only the first two stages overall, namely Exploration and Sensemaking and Collaboration and Delegation. These include P22, P23, and P9, whose trajectories did not progress into Cognitive Disengagement, as well as P2, P18, P27, and P29, who adopted CGM relatively late and therefore may not yet have had sufficient time to develop a later-stage trajectory. Among these participants, P9, P22, and P23 are particularly informative because they show resistance to the proposed three-stage account, whereas the four later adopters cannot yet confirm or contradict the later course of the account.

P11, P13, and P21 (for example, P21: \textit{``Just let it handle things, what do I know, no need to worry about it needlessly''} [2025]) displayed consistently low interpretive engagement across the observed waves. The interviews suggest that this pattern may be associated with their lower educational background, limited digital literacy, and strong deference to technological systems. These cases indicate that the same behavioral endpoint can be reached through at least two distinct pathways: an erosion pathway, in which cognitive disengagement as defined in §2 (a longitudinal decline from earlier active engagement) genuinely occurs, and a stable-low pathway, in which interpretive engagement never rose in the first place and therefore does not constitute disengagement in the strict sense, though it converges on a similar behavioral profile.

The above classification is based on the overall coding results of all five interview waves, rather than solely on the representative quotations shown in the text; the quotations presented in the text are only illustrative examples.

Early attrition and counter-trend cases. Eight participants did not complete the study in full. P30 and P31 withdrew starting from Wave 2; P28 and P35 withdrew starting from Wave 3; P7 withdrew starting from Wave 4; P4 and P14 withdrew starting from Wave 5; P33 missed Wave 2 due to work reasons but participated in all other waves. For those who withdrew and could still be reached (P7, P28, P31, P35), we conducted brief follow-up interviews in 2025 to gain a basic understanding of the changes they experienced after withdrawing. These included: (1) Continued non-adoption of CGM: P7 followed medical advice and did not use CGM due to mild diabetes. (2) Active refusal after a trial: P28 tried CGM early on but refused to continue using it due to safety concerns. P28: "Keeping something stuck in my skin for so long doesn't feel very healthy."(this was still grounded in the doctor's judgment that his condition was relatively mild, making CGM not strictly necessary or acceptable to go without). (3) Life-stage-related withdrawal: P31 and P35 withdrew for reasons unrelated to the system (relocation, family circumstances, worsening health). These cases of withdrawal do not, in themselves, provide information relevant to the trajectory framework; rather, they constitute right-censoring, and we do not draw inferences about cognitive disengagement from them.

We examined four subgroup comparisons, as described below (As shown in Figure 2).

\begin{figure}[htbp]
    \centering
    \includegraphics[width=0.8\textwidth]{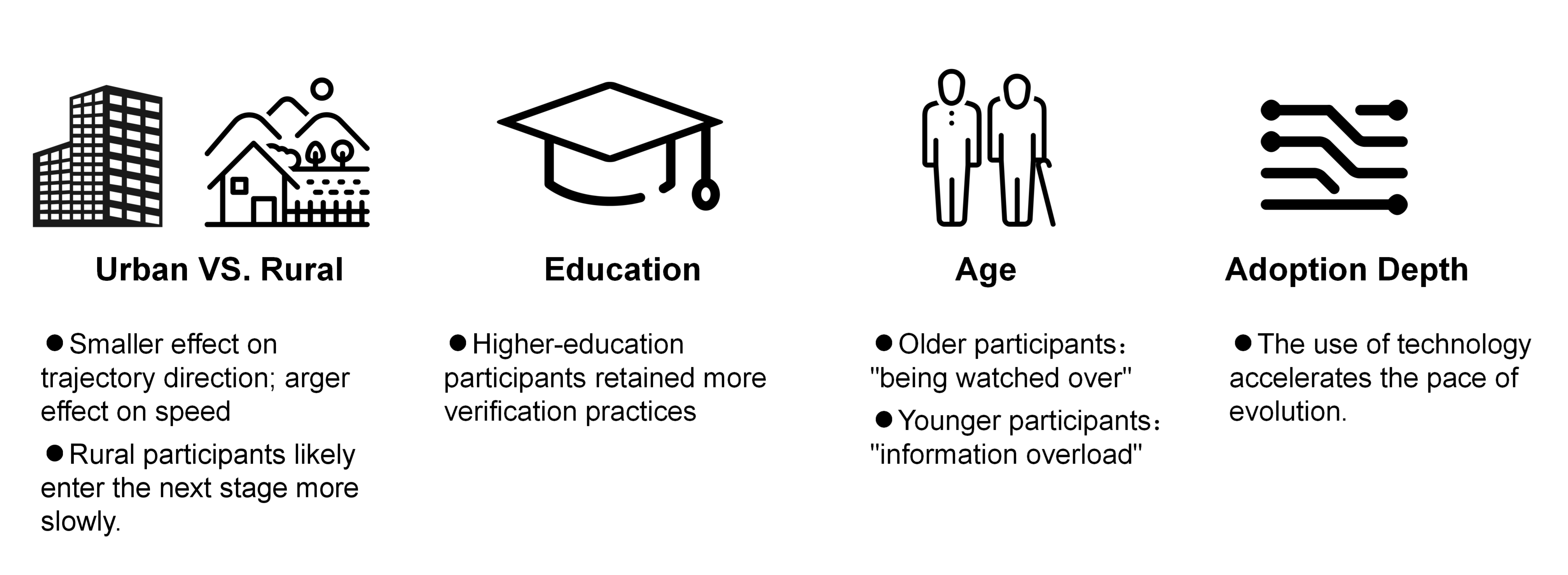}
    \caption{Subgroup analyses by urban/rural area, education, age, and adoption depth}
    \label{fig:figure2}
\end{figure}

Urban-rural comparison. Among those who completed Wave 5, the trajectory patterns of the urban (n=15) and rural (n=14) subgroups were broadly similar overall. However, rural participants generally gained access to new technological systems later, which meant that they entered each of the three stages of cognitive evolution later as well.

Education level. Participants with higher education (bachelor's degree or above; n=4 in the completer group: P10, P12, P22, P23) more often retained partial verification practices at later stages; two of them, P22 and P23, had not entered Stage 3 within the observed window. Participants with lower education (primary school or below; n=17 in the completer group) on average entered Stage 3 earlier, and some skipped Stages 1 and 2 entirely. Higher education was thus associated with slower progression in most cases and, for P22 and P23, with the trajectory's endpoint itself remaining unreached over this eight-year horizon. Given the small subgroup size (n=4) and the absence of statistical controls for confounds such as technology-adoption depth, this pattern should be read as descriptive and hypothesis-generating rather than a tested moderation effect.

Age. Both older participants (age $>$60 at the start of the study; n=6 in the completer group) and younger participants (age $\leq$50 at the start of the study; n=13 in the completer group) displayed these trajectories. Older participants tended to describe the process using phrases like ``the device is watching for me (P13)'' (a stance of letting go), while younger participants tended to describe it using phrases like ``not enough time/too much information (P36)'' (a stance of offloading).

Depth of technology adoption. As described earlier, groups that adopted the technology earlier generally progressed through the three-stage account's trajectory more quickly. In addition, sudden-event cases of technology use (such as P9's AID incident) also prompted users toward reflection.

\section{Discussion}\label{sec:discussion}

Our results show that long-term use of wearable diabetes technologies involved changing forms of use rather than a stable pattern of continued adoption. Specifically, users' cognitive engagement with diabetes-monitoring data changed over long-term use as the systems themselves developed from measurement and visualization toward real-time feedback, prediction, and automated decision-making. We found that many participants moved across three broad stages: self-initiated exploration and sensemaking, collaboration and delegation, and continued system use accompanied by reduced cognitive engagement. These stages represent recurring configurations of system capabilities and user practices rather than fixed chronological boundaries or a universal sequence completed by every participant.

\subsection{Three Stages of Changing Cognitive Engagement}
Our longitudinal analysis identified three broad stages in how users engaged cognitively with diabetes-monitoring systems, described in detail in Section 4.2. These stages broadly corresponded to changes in system capabilities, but they did not constitute a technologically determined or inevitable progression: participants entered them at different times, some did not experience all three, and a small number retained or regained active interpretive practices. The stages describe recurring configurations of technology use, while individual trajectories describe how participants moved through, remained within, or occasionally returned to these configurations over time.

The first stage, exploration and sensemaking, aligns with Mamykina et al.'s \cite{mamykina2017personaldiscovery} model of personal discovery in chronic disease self-management, in which users iterate through feature selection, hypothesis generation, evaluation, and goal specification to establish cause-and-effect relationships between daily behaviors and glucose dynamics. It is also consistent with personal informatics research that treats reflection as a core component of self-tracking \cite{li2010stagebased, mamykina2017personaldiscovery}. Such intensive engagement appears most prominent when continuous data is still novel and system outputs still require users to compare, contextualize, and interpret information themselves.

The second stage, collaboration and delegation, in which real-time feedback and recommendations redistributed part of the interpretive work to the system, aligns with Kabir and Wiese's synthesis of design and implementation factors supporting engagement in personal informatics systems \cite{kabir2023meta}. Delegation remained partial: participants still communicated with clinicians, interpreted data in some cases, and uploaded materials. This stage is therefore better understood as system-supported collaboration than complete withdrawal, extending work on semi-automated tracking \cite{choe2017semi}, where meaningful participation persisted alongside automation.

The third stage, cognitive disengagement, in which system operation moved further into the background while engagement declined, maps onto two categories in Kabir and Wiese's taxonomy \cite{kabir2023meta}: fluctuating motivations (loss of novelty and perceived urgency) and system-caused harm (fatigue, distraction, and obsession from repetitive prompts). Unlike the lapsing or abandonment described in lived-informatics accounts \cite{rooksby2014lived,epstein2015lived}, device use and data collection continued even as self-initiated interpretation declined. This extends Choe et al.'s \cite{choe2014understanding} account of difficulties from tracking excessive information by showing how increasingly directive systems can reduce users' perceived need to interpret that information themselves.

Participants who did not follow the full account refine rather than weaken it. Some experienced only the first two stages, showing that increasingly capable systems do not inevitably reduce engagement. Others remained in a low-engagement configuration throughout, indicating that low engagement at one point should not automatically be treated as disengagement, which requires a longitudinal reduction from earlier active practice. Cases of resumed interpretation after system failure or reflection show that movement into the third stage need not be irreversible.

\textbf{Continued Use with Cognitive Disengagement:} The central finding of our study is a previously under-theorized user state: \emph{continued use with cognitive disengagement}. The device is still worn, the data pipeline still operates, and the user has not behaviorally abandoned the system, yet active interpretation, model-building, questioning, and learning occur less frequently, with users engaging with data mainly when prompted by an alarm or recommendation.

This state differs from adjacent concepts as distinguished in Section 2.1. It is not abandonment: existing abandonment research \cite{epstein2016beyond,clawson2015noaccount} assumes disengagement is observable in usage logs, whereas our state is behaviorally invisible, with users remaining active on every external indicator. It is not automation complacency, which concerns insufficient monitoring of highly reliable automation in supervisory-control settings \cite{parasuraman2010complacency}; our finding concerns a broader longitudinal reduction in self-initiated interpretation not limited to monitoring failures. It is also not simply deskilling: Budzyn et al.\ \cite{budzyn2025deskilling} document a measurable decline in adenoma detection after AI assistance is removed, the local degradation of one professional skill, whereas what we observe is more diffuse, reduced engagement across observation, comparison, inference, modeling, and questioning.

This places \emph{continued use with cognitive disengagement} in a specific gap: distinct from abandonment, broader than complacency, and different in scope from locally measured deskilling. It connects to but is not reducible to theories of \emph{transactive memory with technology} \cite{sparrow2011google} and the \emph{extended mind} \cite{clark1998extended,heersmink2017extended}, which predict that delegating cognition to artifacts is normal and often beneficial but do not predict that such delegation will, in long-term use, reduce how often users exercise their own interpretive practices or how prepared they remain to resume them.

This concept has direct implications for three communities. For personal informatics, the reflection stage in Li et al.'s model \cite{li2010stagebased} is not merely skippable but may become progressively less frequent over long-term use, so system value should be measured by depth of engagement rather than retention or usage rate alone. For automation-trust research, appropriately calibrated trust \cite{lee2004trust} can coexist with a longitudinal reduction in self-initiated verification, suggesting trust calibration alone does not capture how engagement changes over extended use. For ubiquitous computing, it shows that Weiser's \emph{disappearance} \cite{weiser1991computer} assumes continued presence equals continued benefit, whereas long-term use can decouple continued operation from continued interpretive participation, so disappearance may be accompanied by reduced interpretation rather than only calmness. This may also be relevant to other systems that continuously sense, interpret, and respond in the background, including smart-home automation, driver-assistance systems, and automated medication support, though applicability beyond diabetes technologies requires further investigation.

\subsection{The Cost of Cognitive Disengagement}
Cognitive disengagement cannot be understood from the user side alone; the design side must be examined in parallel. Theoretically, this aligns with mediation theory in philosophy of technology, which holds that artifacts co-shape both action and perception \cite{verbeek2005what}: design choices do not merely facilitate use but actively reorganize what users can see, do, and reason about. Our interviews with clinicians, designers, and vendors revealed three mutually reinforcing logics that pre-configure the conditions for disengagement. \textit{Clinical prioritization}: physicians' core expectation is timely intervention rather than support for understanding (Doctor1: ``We mainly look at whether the patient will be in danger\ldots\ why it happened is not the most critical''). When this prioritization is translated directly into design orientation, supporting user understanding can be displayed from the core objectives. \textit{Information compression in design practice}: designers treat ``reduce user burden'' as the primary goal and compress complex data into simple signals (Designer2: ``We try to turn complex information into simple signals''). This lowers short-term comprehension thresholds but simultaneously closes off users' access to raw data structures and the chance to develop their own interpretive frameworks. \textit{Engineering automation}: vendors describe their direction as ``minimize user operations'' (Vendor2), which equates user participation with user burden and ignores that participation is also how understanding is built. Each logic is locally rational, yet their confluence produces a collective consequence no individual stakeholder explicitly intends: the system is built as a \emph{judgment substitute} rather than an \emph{understanding scaffold}. The deeper implication is a structural design blind spot: when each logic operates within its own frame, no frame includes the long-term maintenance of users' interpretive practices and readiness to resume them.

Although we found these design logics in diabetes care, the same underlying problem may arise in fields with different stakeholders and organizations. More broadly, systems are often designed to work well and require less user effort, while helping users maintain their understanding over time is not treated as an explicit design goal. As ubiquitous systems take over more of the work of interpreting data, users may continue using them but interpret and question their outputs less often on their own. This does not mean that other fields will follow the same three stages found in our study. Instead, our findings raise a broader question for ubiquitous computing research: when systems require less interaction, do users also become less likely to actively understand and question their outputs?

Weiser's calm-technology vision \cite{weiser1996calm} silently assumes that as the technology recedes into the background, the user's competence and judgment remain intact in the foreground. Our data suggest an additional complication: as technology recedes, users' interpretive practices may also become less frequent. We describe this outcome as \emph{reduced independent interpretive engagement}: users sustain behavioral health-management practices under continuous system support while becoming less involved in independently interpreting their own data. Reduced interpretive engagement is invisible while the system is functioning; it becomes \emph{potentially} consequential only when system protection is interrupted (failure, misconfiguration, perceptual barriers). The case of P15 makes this exposure mechanism observable in a low-stakes form: a configuration error muted alarms, P15 had hearing impairment, and her trajectory had shifted toward equating ``no alarm'' with ``normal''---all three preconditions for a \emph{double vulnerability}, in which system support is unavailable at the same time that independent checking has become less routine, were directly evidenced in our interviews. This methodological caution matters because cross-sectional research cannot in principle observe the cumulative attrition that double vulnerability presupposes; longitudinal qualitative work can surface the mechanism but cannot, by itself, quantify how often it leads to harm. From the human-factors perspective, the pattern is consistent with Parasuraman and Riley's warning about misuse \cite{parasuraman1997humans}, but in a more insidious form: the failure mode of interest is not a single bad trust judgment but the declining frequency of independent judgment and reduced readiness to resume it after years of normal use. We treat this as a qualitatively supported hypothesis that should be tested in controlled engineering, clinical, and behavioral studies.

This disengagement materializes in two structural consequences, each a face of the same underlying process. (1) \textit{Health risks}: independent perception and judgment of bodily state weaken in tandem with delegation to the system. Some users equate ``no alarm'' with ``normal,'' a judgment grounded in trust rather than understanding. When the system fails to alarm for any reason, users may be exposed to the double-vulnerable state described above. We are careful here to distinguish two epistemic claims: that participants have shifted to ``no-alarm-equals-normal'' reasoning is directly evidenced in our interviews; that this shift translates into adverse outcomes \emph{rate} is something our design cannot estimate. Existing AID research \cite{barnard2015psychosocial,lawton2020clinician} documents trust tensions at the introduction stage; our data show how the early protective tension is dissolved by accumulated normal-use experience, and we identify the resulting double-vulnerable state as a target for future quantitative and engineering investigation. (2) \textit{Burden transfer}: systems do not actually reduce total cognitive burden; they transform it. The burden of \emph{understanding complex data} is converted into the burden of \emph{continuously responding to system prompts}, an interaction-design problem long studied in the notification literature \cite{pielot2014situ} and prefigured in Coiera's analysis of communication overload in healthcare \cite{coiera2003communication}. The former is deep, constructive cognitive work whose by-product is accumulated understanding; the latter is shallow, reactive looping whose by-product is fatigue and further withdrawal. Crucially, the new burden is hidden: it does not present as ``too hard to understand'' but as ``annoying,'' ``tiring,'' ``can't be bothered''---emotions that may lead users to participate less rather than sustain interpretive practices, potentially reinforcing further disengagement. The two consequences share a common condition: when self-initiated interpretation becomes less frequent, users may be less prepared to independently evaluate bodily state and the salience of system prompts. These consequences therefore should not be addressed only as isolated failures; they also require attention to how system design sustains users' cognitive participation over time.

\subsection{Design Implications: Preserving Interpretive Engagement Under Automation}
We propose shifting system design from \emph{a machine that substitutes for judgment} to automation that reduces unnecessary cognitive work while preserving users' opportunities to understand, contextualize, and question outputs. This is not a rejection of automation but a redesign of its relationship to understanding. Preserving interpretive engagement does not require continuously examining all data or reproducing the system's calculations; it means keeping relevant evidence, uncertainty, and context accessible when users need to evaluate consequential outputs. Since this study did not experimentally test interface interventions, the implications below are mechanism-informed directions, not demonstrated solutions.

\textit{1. Provide layered explanations at consequential moments.} Participants such as P19 and P14 shifted from examining why glucose changed to directly responding to system conclusions (Section 4.2.2). As outputs grew more directive, the evidence needed to evaluate them became less central to routine interaction, though this does not show explanations alone prevent disengagement or automation bias, only that layering can preserve access to judgment bases without requiring continuous processing.

Beyond alerts, systems should communicate glucose direction and rate, the data window behind an alert, prediction uncertainty, and contributing readings or patterns, without unsupported causal claims. An initial message can state the condition and recommended action, with trend, data, and limitations available on further inspection. This responds jointly to P19 and P14 examining output bases less often, and P17, P24, and P26 describing continuous processing as tiring; presenting all detail by default risks reproducing that burden, while layering may preserve optional access without it, though this needs empirical evaluation.

Explanatory support should remain accessible after adoption rather than withdrawn as use becomes routine. Section 4.4's variation, some participants retaining verification, others still questioning delegated judgment, suggests interpretation stays useful post-adoption, though not that any specific feature reactivates engagement. Periodic summaries highlighting recurring patterns, recommendation changes, or unresolved anomalies could support selective reflection without maximizing interface activity; reduced activity alone should not signal reduced understanding, and such features should stay optional and adjustable.

\textit{2. Automate routine work while preserving contextual contribution.} Automating sensing, storage, and routine calculation reduces unnecessary burden, but removing channels through which users, care partners, and clinicians contribute information the system lacks can restrict meaningful participation. This does not mean manual interaction prevents disengagement; it means retaining ways to add observations, experience, or care-related knowledge.

Food-tracking and data-sharing practices (Sections 4.1.2, 4.2.2) provide boundary cases against simple substitution: P25 and P26 continued entering dietary information, P15 and P17 used shared data to involve family and clinicians, showing partial automation can generate new contribution and collaborative interpretation rather than reduce it. These support retaining contextual and coordination channels, though not that any specific feature sustains engagement universally; the implication is only that automation should not eliminate contribution forms that proved meaningful.

Consistent with semi-automated tracking \cite{choe2017semi}, systems may automate routine capture and calculation while letting users annotate events, report symptoms, correct assumptions, and formulate clinician questions, prioritizing features where contributed information changes interpretation or resolves what automated data cannot. Manual interaction should not be forced when it adds little value, since participants also found repeated prompts burdensome; input without clear purpose could intensify rather than reduce disengagement. The goal is automating low-value burden while preserving agency and contextual knowledge, not maximizing either automation or manual work.

\textit{3. Design for failures in sensing, alerting, and perception.} Existing systems assume sensing, processing, and notification work as intended; designers must also address degraded conditions, incomplete data, configuration errors, inaccessible notifications, and conflicts between output and bodily experience. This does not show reduced interpretation means lost judgment, only that some participants increasingly relied on alerts to determine when action was needed.

P15 and P9 provide direct motivation: P15 assumed no perceived alert meant normal, though her device volume was minimized and hearing impairment blocked some alerts; P9 resumed independent checking after an AID malfunction. Both show protection can fail not just through algorithm or sensor error but when alerts fail to reach the user, a vulnerability arising when attention had become organized around outputs while the depended-on channel became unreliable. The objective is not continuous manual monitoring but making degraded states detectable, with clear routes for verification and recovery.

Systems should show sensor connectivity, data completeness, alarm volume, notification permissions, and delivery status; safety-critical alerts should use redundant modalities (sound, vibration, visual, care-partner communication); periodic checks may catch muted or inaccessible alerts early. Interfaces should avoid implying alert absence means normal condition, offering verification and escalation guidance when symptoms conflict with output or sensing status is uncertain, directly targeting the P15 and P9 failure conditions. Unlike layered explanations, which preserve access during normal operation, failure-oriented design should communicate channel status and next steps when channels fail, keeping status, delivery, and fallback accessible when automation becomes unavailable or hard to perceive.

\textit{Cross-cutting implications for long-term evaluation.} Because the three implications address different mechanisms, they warrant distinct evaluation rather than one engagement measure. Continued device use, data continuity, interface openings, and alert responsiveness are insufficient indicators, given the heterogeneous patterns observed (retained verification, re-engagement after failure, late adoption, consistently low engagement). Evaluation should examine, respectively, whether users can access and understand consequential outputs' basis and limitations; whether contextual-input and sharing features contribute information or support collaboration without unnecessary burden; and whether users can recognize degraded conditions and initiate verification or escalation.

These assessments should not assume every user must independently perform all calculations or continuously analyze data, nor infer interpretive needs solely from age, education, adoption time, or interface-use frequency. Evaluation should examine whether automation remains understandable, contestable, and recoverable when questioned, and test whether proposed features achieve their intended effects, since the findings identify problems to address without establishing that any strategy will prevent disengagement.

Together, these implications call not for less automation but for automation that reduces unnecessary cognitive work without hiding system reasoning, preserves contextual participation, and supports users when routine protection fails, distinguishing access to reasoning during normal operation, channels for human contribution, and detection and recovery when protection degrades.

\subsection{Generative Implications for Broader Ubiquitous Computing Design}

While our observations are grounded in longitudinal diabetes management, the core phenomenon we identify, continued use with cognitive disengagement, is not unique to chronic disease care. CGMs and AID systems represent a paradigmatic micro-cosmos of modern Ubiquitous Computing, embodying structural characteristics increasingly common across ubiquitous environments: continuous background sensing, algorithmic prediction, and progressive delegation of decision-making. Analyzing these technologies over an eight-year horizon offers a generalizable conceptual lens and generative hypotheses for the broader UbiComp community.

\subsubsection{Revisiting the Calm Technology Ideal: Disappearance vs. Cognitive Attrition}
A foundational tenet of ubiquitous computing, from Weiser's vision of calm technology \cite{weiser1991computer,weiser1996calm}, holds that systems should recede into the background, enabling effortless operation. Our findings qualify this: technological disappearance does not guarantee cognitive empowerment, and may instead mask a silent decay in independent interpretive practice.

Classical UbiComp frameworks often treat low interface interaction as a success metric, indicating seamless automation and high trust. Our analysis shows that as systems shift from decision-support (visualizing trends) to decision-substitution (issuing directives or automated actions), "calmness" can decouple from "informed agency": users may develop a heuristic equating the absence of prompts with safety ("no alarm equals normal"). This reveals a design tension in pervasive systems: \emph{the mechanisms that reduce short-term interaction friction can systematically erode users' readiness to inspect, understand, and override automated systems when boundaries fail.}

\subsubsection{Generative Hypotheses Across Pervasive and Agentic System Classes}
Our findings are rooted specifically in chronic disease care, and the multi-stage trajectories observed should not be over-generalized as laws governing all pervasive systems. However, the underlying structural drivers are actively shaping interaction paradigms across broader UbiComp domains. We propose continued use with cognitive disengagement not as established fact elsewhere but as a sensitizing framework for two emerging system classes:

\begin{itemize}
    \item \textbf{Automated Driving and Driver-Assistance Systems (ADAS):} In Level 2+ autonomous driving, vehicle systems handle routine steering, acceleration, and lane-keeping \cite{nidamanuri2021progressive}. As our participants equated the absence of CGM alarms with safety, drivers experiencing years of reliable automated assistance may undergo a similar drift in situational awareness, with seamless automation degrading mental models of edge cases and reducing readiness for manual takeover during sudden degradation.
    \item \textbf{Personal Productivity and Emerging Agentic AI Systems:} Intelligent email summarizers, scheduling assistants, and autonomous AI agents compress complex decision spaces into single-click approvals or automated executions \cite{faulring2010agent,chen2026not,liu2025supporting}. As such systems evolve from decision support to judgment substitution, users face a structural transfer of cognitive load, from deep, constructive sensemaking to shallow, reactive prompt verification.
\end{itemize}

These cross-domain connections are theoretical extrapolations and generative hypotheses, not empirically proven facts outside clinical contexts. Whether technological "disappearance" \cite{weiser1991computer} leads to calm empowerment or cognitive impoverishment in these adjacent domains remains an urgent empirical question. By theorizing continued use with cognitive disengagement, we urge IMWUT researchers to evaluate long-term system value not by behavioral retention, usage frequency, or algorithmic accuracy alone, but by the longitudinal preservation of human interpretive agency.

\subsection{Limitations and Future Work}

\textbf{Limitations}: (i) Our 37 participants come from a specific region; while diverse in urban/rural balance, education, and occupation, they are not statistically representative. 
The stages and trajectories we describe are process-level mechanisms in a particular setting, not population-level estimates; its generalizability to other healthcare systems and cultural contexts requires further study, especially in low- and middle-income settings where CGM/AID adoption is still highly unequal~\cite{find2024cgm}. Frameworks for assessing technology adoption such as NASSS~\cite{greenhalgh2017beyond} could provide a useful comparative scaffold for replication studies. (ii) Our design implications Section 5.3 are inferred from longitudinal interview analysis, not validated through design intervention. They represent reasoned design directions, not verified solutions, and RQ2 is therefore answered at the level of design problems and directions rather than tested solutions. Future research should prototype, deploy, and longitudinally evaluate these directions.

\textbf{Future directions}: (i) Replicate the longitudinal design across cultural and healthcare-system contexts to test the cross-context stability of the cognitive-disengagement pattern. (ii) Develop measurement instruments for ``depth of cognitive engagement'' that can be tracked in both research and operational systems. (iii) Design and evaluate concrete interfaces and interaction mechanisms that preserve interpretive engagement under automation, with controlled studies of their effects on engagement depth and long-term health autonomy. (iv) Extend the framework to other long-term high-dependence everyday computing systems beyond health, to test the boundary conditions of cognitive disengagement as a general phenomenon. (v) Section 4.4 provides an initial, 
within-cohort comparison of users at different technology-adoption depths; a matched quasi-experimental cohort design would extend this descriptive pattern with stronger causal leverage, more cleanly separating (a) long-use, (b) technology, and (c) life-course contributions to cognitive disengagement.

\section{Conclusion}\label{sec:conclusion}
Through an eight-year, five-wave longitudinal study of 37 patients with diabetes across three sociotechnical generations of diabetes  computing systems, we identify a previously under-theorized user state---\emph{continued use with cognitive disengagement}---in which the device remains worn and the data pipeline keeps running while active interpretation, model-building, and questioning have all withdrawn. This state is pre-configured by the confluence of clinical, designerly, and engineering logics that build systems to \emph{substitute} for, rather than \emph{scaffold}, user understanding, and it materializes as independent health judgment occurring less frequently, and a transfer of burden from understanding to responding.

\newpage
\bibliographystyle{ACM-Reference-Format}
\bibliography{sample-base}

\onecolumn

\appendix

% =========================================================
% A. Primary Participant Information
% =========================================================

\section{Primary Participant Information}

\begingroup
\scriptsize
\renewcommand{\arraystretch}{1.0}
\setlength{\tabcolsep}{2pt}

\refstepcounter{table}
\label{tab:primary_participants}

\begin{center}
\textbf{Table \thetable. Primary Participant Information}

\vspace{4pt}

\begin{tabular}{@{}
p{0.035\textwidth}
p{0.040\textwidth}
p{0.055\textwidth}
p{0.040\textwidth}
p{0.135\textwidth}
p{0.140\textwidth}
ccccc
@{}}

\toprule
\textbf{ID} &
\textbf{Gen.} &
\textbf{Resid.} &
\textbf{Age} &
\textbf{Education} &
\textbf{Occupation} &
\shortstack{\textbf{2017}\\\textbf{Att.}} &
\shortstack{\textbf{2019}\\\textbf{Att.}} &
\shortstack{\textbf{2021}\\\textbf{Att.}} &
\shortstack{\textbf{2023}\\\textbf{Att.}} &
\shortstack{\textbf{2025}\\\textbf{Att.}} \\
\midrule

P1  & M & Rural & 55 & Junior High School & Service Staff   & Yes & Yes & Yes & Yes & Yes \\
P2  & M & Rural & 56 & Primary School     & Skilled Laborer & Yes & Yes & Yes & Yes & Yes \\
P3  & F & Urban & 49 & Semi-literate      & Cashier         & Yes & Yes & Yes & Yes & Yes \\
P4  & F & Rural & 39 & Junior High School & Courier         & Yes & Yes & Yes & Yes & No  \\
P5  & M & Rural & 44 & Junior High School & Service Staff   & Yes & Yes & Yes & Yes & Yes \\
P6  & F & Urban & 47 & Junior High School & Retail Worker   & Yes & Yes & Yes & Yes & Yes \\
P7  & F & Urban & 57 & Primary School     & Bus Driver      & Yes & Yes & Yes & No  & No  \\
P8  & F & Rural & 59 & Primary School     & Retail Worker   & Yes & Yes & Yes & Yes & Yes \\
P9  & M & Rural & 64 & High School        & Unemployed      & Yes & Yes & Yes & Yes & Yes \\
P10 & M & Urban & 65 & Bachelor's Degree  & Unemployed      & Yes & Yes & Yes & Yes & Yes \\
P11 & M & Urban & 67 & Primary School     & Cleaner         & Yes & Yes & Yes & Yes & Yes \\
P12 & F & Rural & 64 & Bachelor's Degree  & Teacher         & Yes & Yes & Yes & Yes & Yes \\
P13 & M & Urban & 66 & Primary School     & Unemployed      & Yes & Yes & Yes & Yes & Yes \\
P14 & M & Rural & 70 & High School        & Unemployed      & Yes & Yes & Yes & Yes & No  \\
P15 & F & Rural & 57 & Primary School     & Caregiver       & Yes & Yes & Yes & Yes & Yes \\
P16 & M & Urban & 59 & Primary School     & Skilled Laborer & Yes & Yes & Yes & Yes & Yes \\
P17 & F & Rural & 58 & Primary School     & Farmer          & Yes & Yes & Yes & Yes & Yes \\
P18 & M & Urban & 56 & High School        & Service Staff   & Yes & Yes & Yes & Yes & Yes \\
P19 & F & Rural & 54 & Primary School     & Farmer          & Yes & Yes & Yes & Yes & Yes \\
P20 & F & Rural & 44 & Primary School     & Factory Worker  & Yes & Yes & Yes & Yes & Yes \\
P21 & F & Rural & 46 & Primary School     & Cashier         & Yes & Yes & Yes & Yes & Yes \\
P22 & M & Urban & 43 & Master's Degree    & Teacher         & Yes & Yes & Yes & Yes & Yes \\
P23 & F & Urban & 47 & PhD                & Professor       & Yes & Yes & Yes & Yes & Yes \\
P24 & M & Urban & 39 & Primary School     & Shop Assistant  & Yes & Yes & Yes & Yes & Yes \\
P25 & F & Urban & 41 & Primary School     & Farmer          & Yes & Yes & Yes & Yes & Yes \\
P26 & M & Urban & 42 & Junior High School & Bus Driver      & Yes & Yes & Yes & Yes & Yes \\
P27 & M & Rural & 49 & Primary School     & Farmer          & Yes & Yes & Yes & Yes & Yes \\
P28 & M & Urban & 67 & Primary School     & Unemployed      & Yes & Yes & No  & No  & No  \\
P29 & M & Rural & 68 & High School        & Unemployed      & Yes & Yes & Yes & Yes & Yes \\
P30 & F & Urban & 59 & Semi-literate      & Street Vendor   & Yes & No  & No  & No  & No  \\
P31 & F & Rural & 53 & Primary School     & Cashier         & Yes & No  & No  & No  & No  \\
P32 & M & Urban & 47 & Primary School     & Hairdresser     & Yes & Yes & Yes & Yes & Yes \\
P33 & M & Rural & 49 & Junior High School & Farmer          & Yes & No  & Yes & Yes & Yes \\
P34 & M & Urban & 54 & Primary School     & Kitchen Worker  & Yes & Yes & Yes & Yes & Yes \\
P35 & F & Rural & 55 & Semi-literate      & Laundry Worker  & Yes & Yes & No  & No  & No  \\
P36 & M & Urban & 45 & Primary School     & Cleaner         & Yes & Yes & Yes & Yes & Yes \\
P37 & M & Rural & 51 & Junior High School & Cleaner         & Yes & Yes & Yes & Yes & Yes \\

\bottomrule
\end{tabular}
\end{center}

\endgroup

% =========================================================
% B. Practitioner Information
% =========================================================

\clearpage
\section{Practitioner Information}

\begingroup
\small
\renewcommand{\arraystretch}{1.05}
\setlength{\tabcolsep}{6pt}

\refstepcounter{table}
\label{tab:practitioners}

\begin{center}
\textbf{Table \thetable. Practitioner Information}

\vspace{4pt}

\begin{tabular}{@{}
p{0.12\textwidth}
p{0.14\textwidth}
p{0.10\textwidth}
p{0.14\textwidth}
p{0.28\textwidth}
@{}}

\toprule
\textbf{ID} &
\textbf{Role} &
\textbf{Gender} &
\textbf{Residence} &
\textbf{Age / Education} \\
\midrule

Doctor1   & Doctor   & M & Rural & 45 / PhD \\
Doctor2   & Doctor   & F & Urban & 44 / PhD \\
Doctor3   & Doctor   & F & Urban & 44 / PhD \\
Doctor4   & Doctor   & M & Urban & 37 / PhD \\
Doctor5   & Doctor   & F & Rural & 39 / PhD \\
Designer1 & Designer & M & Rural & 26 / Bachelor's Degree \\
Designer2 & Designer & F & Urban & 29 / Bachelor's Degree \\
Designer3 & Designer & M & Urban & 34 / Bachelor's Degree \\
Designer4 & Designer & F & Urban & 33 / Bachelor's Degree \\
Designer5 & Designer & M & Rural & 37 / Bachelor's Degree \\
Vendor1   & Vendor   & M & Urban & 45 / High School \\
Vendor2   & Vendor   & F & Rural & 47 / Bachelor's Degree \\
Vendor3   & Vendor   & F & Rural & 51 / Bachelor's Degree \\
Vendor4   & Vendor   & M & Urban & 48 / Bachelor's Degree \\
Vendor5   & Vendor   & F & Rural & 39 / Master's Degree \\

\bottomrule
\end{tabular}
\end{center}

\endgroup

% =========================================================
% C. Interview Guide for Primary Participants
% =========================================================

\clearpage
\section{Interview Guide for Primary Participants}

\begingroup
\scriptsize
\renewcommand{\arraystretch}{1.10}
\setlength{\tabcolsep}{3pt}

\refstepcounter{table}
\label{tab:interview_primary}

\begin{center}
\textbf{Table \thetable. Interview Guide for Primary Participants}

\vspace{4pt}

\begin{tabular}{@{}
p{0.16\textwidth}
p{0.34\textwidth}
p{0.42\textwidth}
@{}}

\toprule
\textbf{Theme} &
\textbf{Core Question} &
\textbf{Possible Probes (Non-leading)} \\
\midrule

Initial Usage Context
& Do you remember when you first started using this device? What was the situation like at that time?
& Who suggested or prompted you to start using it? Was there anything specific that left a strong impression on you? \\

Most Recent Use
& When was the last time you noticed or interacted with this device or system? What happened at that time?
& Did you check it yourself, or did it prompt you? What were you doing at that moment? \\

Everyday Use Episodes
& In your daily life, are there certain moments when you regularly check this system? Can you give a specific example?
& For example, during meals, exercise, or rest? Are there times when you would not check it? \\

No-Alert Situations
& If the system does not provide any alerts for a period of time, what do you usually do?
& Do you check it proactively, or do you generally not pay attention to it? Are there any exceptions? \\

Triggered Event Handling
& When the system alerts you about something, how do you usually respond? Can you describe a recent example?
& Did you handle it immediately, or did you delay it? Are there situations where you did not act on it? \\

Decision-Making Process (Behavior)
& In the example you just mentioned, how did you decide what to do?
& Did you refer to anything else, such as bodily feelings, prior experience, or other people, or did you mainly rely on the system? \\

Body vs.\ System
& Have you ever encountered situations where your bodily feelings did not match what the system showed? How did you handle that?
& Did you try to verify it, or did you follow the system directly? \\

Long-Term Changes (Behavior)
& Compared with when you first started using it, has the way you use this system changed?
& For example, frequency of checking, usage scenarios, or whether you still deliberately spend time checking it? \\

Attention Allocation
& How much attention does this system take in your daily life?
& Do you think about it often, or only notice it when it alerts you? \\

Information Processing Burden
& During use, have you ever felt that you needed to process a lot of information or make many decisions?
& When does this usually happen? How do you cope with it? Are there times when you choose to ignore it? \\

Ignoring and Coping
& Have there been situations where the system prompted you but you did not follow it?
& Why did you not act at that time? What happened afterward? \\

Data-Related Understanding
& Have you ever thought about how these data are recorded or transmitted to your phone?
& Do you check or adjust related settings, or do you generally not look into them? \\

Role of the System in Life
& In your daily life, what does this system feel like to you?
& For example, a reminder tool, something that records information, or something else? \\

Extreme or Memorable Experiences
& Have you had any particularly memorable experience while using this system?
& What happened at that time? How did you respond? \\

Ideal Improvements (Open)
& If you could change it, in what ways would you want it to be different?
& Are there aspects that you feel are not quite appropriate in its current form? \\

\bottomrule
\end{tabular}
\end{center}

\endgroup

% =========================================================
% D. Interview Guide for Doctors, Designers, and Vendors
% =========================================================

\clearpage
\section{Interview Guide for Doctors, Designers, and Vendors}

\begingroup
\scriptsize
\renewcommand{\arraystretch}{1.10}
\setlength{\tabcolsep}{3pt}

\refstepcounter{table}
\label{tab:interview_practitioners}

\begin{center}
\textbf{Table \thetable. Interview Guide for Doctors, Designers, and Vendors}

\vspace{4pt}

\begin{tabular}{@{}
p{0.16\textwidth}
p{0.34\textwidth}
p{0.42\textwidth}
@{}}

\toprule
\textbf{Theme} &
\textbf{Core Question} &
\textbf{Possible Probes (Non-leading)} \\
\midrule

Typical Usage Scenarios
& In your experience, in what situations do users typically use this system?
& Are there any usage scenarios that left a strong impression on you? \\

Key Design Decisions
& During design or implementation, were there any important decisions that were repeatedly discussed?
& What trade-offs were considered at that time? Why was the final decision made this way? \\

Data Presentation Choices
& How was the current way of presenting information in the system determined?
& Were alternative presentation approaches considered? Why were they not adopted? \\

Source of Alert Mechanisms
& Under what conditions does the system issue alerts? How were these rules established?
& Are they based on experience, clinical standards, or testing? \\

Assumptions About User Actions
& In your expectation, how should users respond after receiving this information?
& Are there behaviors that you would prefer users not to do? \\

Boundaries of Automation (Case-Based)
& Are there situations where you would consider letting the system handle things automatically rather than relying on users?
& Why do you think these tasks are suitable for automation? \\

Observations of User Practice
& Have you observed how users actually use this system?
& Are there discrepancies between actual usage and your original expectations? \\

Information Selection
& During design, were there types of information that you decided not to present to users?
& Why were they not presented? What concerns were involved? \\

Feedback on Usage Problems
& What are the most common issues reported by users?
& Are there problems that you find particularly difficult to solve? \\

Multi-Party Information Distribution
& How is information distributed among users, family members, and doctors?
& Have there been situations where information was inconsistent or conflicted? \\

Risk Handling Cases
& Have there been cases where system judgment or usage led to problems?
& How were they handled at the time? Were there subsequent adjustments? \\

Long-Term Usage Changes
& Have you observed changes in user behavior after prolonged use?
& For example, reduced usage or changes in usage patterns? \\

Practical Constraints
& In real-world development or clinical settings, what factors constrain system design?
& Technical limitations, responsibility, time, regulations, and so on. \\

Ideal System (Open)
& If there were no constraints, what would you want such systems to become in the future?
& Are there features you have always wanted to implement but could not? \\

\bottomrule
\end{tabular}
\end{center}

\endgroup

\section{Detailed Study Design Rationale}\label{app:studydesign}

This appendix provides additional rationale and procedural details for the longitudinal study design summarized in \S\ref{sec:method}.

\paragraph{Rationale for five waves and two-year intervals.}
The two-year interval between waves was chosen to balance three considerations: (1)~allowing enough time for meaningful changes in technology, interpretive practice, and usage patterns to accumulate between observations; (2)~keeping the interval short enough to minimize retrospective bias when participants recall events since the previous interview; and (3)~practical feasibility of maintaining a cohort relationship over the full study period. The five-wave structure (2017, 2019, 2021, 2023, 2025) was designed to span what we anticipated, based on early-wave data, to be the full arc from early adoption through mature use. In retrospect, the eight-year span proved sufficient to observe the complete trajectory from active engagement to cognitive withdrawal in the majority of participants, though some late-adopting participants may require additional follow-up to determine whether their trajectories ultimately converge with the dominant pattern.
The eight-year span was necessary because the trajectory we report, the gradual shift from exploration and sensemaking, through collaboration and delegation, to cognitive disengagement (§4.2), together with the accompanying technological changes across three generations (fingerstick meters, CGM, and predictive/AID systems), took most participants this long to unfold. A shorter window would have captured only the early stages and never observed the trajectory's endpoint.

\paragraph{Rationale for the dual-group design.}
The primary participant group (patients) provides within-subject longitudinal data on the user side of the human--system relationship. The supplementary participant group (clinicians, designers, vendors) was introduced as a single-wave complement in 2025 specifically to illuminate the \emph{design-side} logics that pre-configure user trajectories. We chose to interview practitioners only in the final wave for two reasons: (1)~the user-side trajectories needed to be sufficiently developed before we could formulate precise questions about design intent, institutional logic, and observed discrepancies between design assumptions and real-world use; and (2)~interviewing practitioners at earlier waves would have risked anchoring our analysis to their explanations rather than allowing user-side patterns to emerge inductively. The practitioner data are therefore used interpretively rather than as an independent longitudinal dataset.

\paragraph{Recruitment and sampling strategy.}
Primary participants (P1--P37) were recruited through a combination of in-situ contact at community health centers and diabetes clinics, and subsequent snowball sampling. We deliberately sought maximum variation \cite{patton1999enhancing} along four dimensions: urban/rural residence, education level, age, and technology adoption history. The initial recruitment (2017, Wave~1) targeted 40 participants to allow for anticipated attrition; 37 completed the Wave~1 interview. Practitioner participants ($n=15$) were recruited through purposive sampling in 2025, with deliberate balance across the three professional roles (5 clinicians, 5 designers, 5 vendors) to prevent any single stakeholder perspective from dominating the design-side analysis. Within each role, we sought variation in institutional setting (public hospital vs.\ private clinic; large manufacturer vs.\ startup; urban vs.\ rural practice) to capture the range of design logics operating across the ecosystem.

\paragraph{Wave-by-wave technology context.}
The five waves correspond to distinct technology contexts experienced by the cohort:
\begin{itemize}
    \item \textbf{Wave~1 (2017)}: Most participants used finger-prick glucometers; a small number had recently begun CGM trials. No participant used predictive systems or AID.
    \item \textbf{Wave~2 (2019)}: CGM adoption had become the majority practice; smartphone connectivity and basic trend visualization were standard. A small number of participants had begun using threshold alarms.
    \item \textbf{Wave~3 (2021)}: CGM with real-time alerts, nighttime hypoglycemia alarms, and dietary/exercise integration was the dominant configuration. Remote cloud-based monitoring by family members or clinicians had emerged.
    \item \textbf{Wave~4 (2023)}: Predictive systems (glucose event forecasting) had entered use for a subset of participants. AID systems were in early trial among the most technologically engaged participants.
    \item \textbf{Wave~5 (2025)}: Predictive systems were available to the majority of the cohort. Multi-device integration (CGM + smartwatch + insulin pump) had become the most advanced configuration in the study.
\end{itemize}

This wave-by-wave technology mapping is essential for interpreting the trajectory analysis, because it shows that the three sociotechnical periods (glucometer-era, CGM-era, predictive/AID-era) were not experienced simultaneously by all participants; the within-cohort variation in adoption timing provides important analytical leverage for separating technology effects from time effects, as discussed in \S4.4. The decision to pair patient interviews with practitioner perspectives is further motivated by broader work on calibrating user expectations of imperfect AI systems, supporting cognitively impaired or aging users with AI assistance, and the expanding clinical role of AI in health and medicine, all of which underscore the value of examining user- and design-side reasoning together \cite{kocielnik2019will,pollack2005intelligent,rajpurkar2022aimedicine,chen2026between}.

\clearpage
\section{Detailed Data-Collection Procedures}\label{app:datacollection}

This appendix provides the full procedural details of the data collection summarized in \S\ref{sec:method}.

\paragraph{Interview logistics and setting.}
All interviews were conducted one-on-one, in quiet, private settings chosen by the participant (typically their home, a community health center meeting room, or a private café room). Each interview lasted 50--73 minutes (mean: 62 min across all waves). All interviews were audio-recorded with the participant's informed consent and transcribed verbatim by a professional transcription service within 72 hours. Transcripts were then verified against the recordings by the interviewing researcher, with particular attention to technical terminology (device names, medical terms) and dialectal expressions.

\paragraph{Longitudinal consistency protocol.}
To ensure comparability across waves, the same interview guide was used at all five time points, with minor additions to capture technology-generation-specific questions (e.g., AID-related questions were added in Wave~4 for participants who had adopted these systems). The interview probes concerning alert frequency and response (see Appendix C) were informed by clinical research on CGM alarm-threshold configuration and its association with glucose outcomes \cite{lin2020alarm}. Each interview began with a brief review of the participant's current device setup and any changes since the previous wave, followed by the four structured phases described in the main text. The interviewing researcher read the participant's prior-wave transcript before each interview to enable informed probing and follow-up on previously reported experiences. This pre-reading was considered essential for longitudinal depth: it allowed the researcher to ask, for example, ``Last time you mentioned you still cross-checked the system against your own feeling---do you still do that?'' rather than generic questions. All three researchers (R1, R2, R3) conducted interviews across all waves; each participant was assigned a primary interviewer who conducted at least three of their five interviews, with the remaining interviews conducted by one of the other two researchers to reduce interviewer bias.

\paragraph{Interviewer conduct.}
Researchers were trained to use open, non-evaluative language throughout. Specific protocols included: (1)~never using the word ``correct'' or ``incorrect'' in response to participants' descriptions of their usage practices; (2)~probing for concrete episodes (``Can you describe the most recent time this happened?'') rather than general attitudes (``How do you feel about the system?''); (3)~using silence and neutral prompts (``Can you tell me more about that?'') to encourage elaboration without shaping content; and (4)~pausing the interview and offering breaks whenever participants showed signs of discomfort, fatigue, or emotional distress. Three participants (P8, P15, P17) required at least one pause during interviews in later waves when discussing health-scare episodes.

\paragraph{Practitioner interviews.}
Interviews with doctors, designers, and vendors were conducted only in 2025, using a parallel semi-structured protocol. These interviews were designed to capture the practical reasoning behind system design and clinical deployment decisions rather than individual user experiences. Each practitioner interview lasted 45--65 minutes (mean: 54 min) and followed a similar format: opening with the participant's professional role and relationship to CGM/AID systems, then moving through four thematic areas: (1)~observations of actual user behavior in clinical or product contexts; (2)~specific design or clinical trade-offs the participant had faced (risk control vs.\ user comprehension, information density vs.\ simplicity, automation vs.\ user involvement); (3)~the practical logics or institutional pressures generating current system forms; and (4)~reflections on gaps between design intent and observed user practice. The full interview guide is provided in Appendix~D.

\paragraph{Participant retention strategy.}
Maintaining a cohort over eight years required deliberate retention effort. We used three strategies: (1)~annual check-in calls (non-interview) between waves to maintain rapport and update contact information; (2)~flexible scheduling that accommodated participants' seasonal and health-related constraints; and (3)~a modest compensation per interview (50 RMB, approximately \$7 USD at the time of each wave), kept deliberately small to avoid incentivizing participation in the absence of genuine willingness. Despite these efforts, 8 of the original 37 participants did not complete all five waves; attrition patterns are analyzed in \S4.4.

\paragraph{Data management and security.}
Audio recordings were stored on encrypted drives accessible only to the three coding researchers. Transcripts were de-identified (participant codes P1--P37 and practitioner codes Doctor1--Doctor5, Designer1--Designer5, Vendor1--Vendor5) immediately after verification. No real names, hospital names, or device brand names appear in the transcripts used for analysis. All data were stored in compliance with the institutional ethics board's data management requirements and will be retained for five years after the final publication, after which they will be securely destroyed.

\section{Detailed Data-Analysis Procedures}

This appendix provides the full procedural details of the data analysis summarized in \S\ref{sec:method}.

\paragraph{Coding unit rationale.}
Following established practice in longitudinal qualitative analysis \cite{saldana2003longitudinal,saldana2009coding}, our basic coding unit was the \emph{participant--wave--episode}: a contiguous narrative segment in which one participant, in one interview wave, described a single concrete usage episode (e.g., a glucose alarm event, a follow-up visit, a morning checking routine). Episodes ranged from one paragraph to roughly two pages of transcript. We chose this unit, rather than the individual utterance, because the theoretical object of interest---interpretive practice---is meaningful only at the episode level, where context, action, and reasoning co-occur. A total of 1,182 episodes were coded across the five waves. This approach is consistent with broader guidance on content-analysis strategy, methodological rigor, and longitudinal design in qualitative health research \cite{hsieh2005three,morse2015critical,young2022series}.

\paragraph{Codebook development and inter-rater reliability.}
The codebook was developed in two pilot rounds (5 participants each) following \cite{macqueen1998codebook,decuir2011developing} and stabilized at 47 codes organized into six families (technology generation, checking practice, interpretive practice, trust posture, alarm handling, data-flow awareness). After codebook stabilization, all three researchers coded the entire dataset independently. Inter-rater agreement was computed on a 20\% random sub-sample (8 participant case files, 246 episodes) using Cohen's $\kappa$ for each pairwise comparison and Fleiss' $\kappa$ for the three-way agreement \cite{mchugh2012interrater,fleiss1971measuring}. Initial pairwise $\kappa$ values ranged from 0.61 to 0.74 (substantial agreement); after dispute resolution, three-way Fleiss' $\kappa$ reached 0.81 on the same sub-sample, which we treat as the reliability floor for the rest of the corpus.

\paragraph{Dispute-resolution protocol.}
Where the three coders disagreed on a code, a structured three-step adjudication was used. (1)~Each researcher articulated their warrant by quoting the specific transcript span and the prior/subsequent waves that informed the assignment. (2)~The three discussed until either consensus was reached or the disagreement was deemed substantive. (3)~If consensus could not be reached, a fourth researcher (R4, who had not coded but was familiar with the dataset) was brought in as adjudicator; if R4 also could not break the tie, the episode was retained but flagged as \emph{interpretively contested} and excluded from claims that depended on it. Of the 246 sub-sample episodes, 31 (12.6\%) reached step~2 and 4 (1.6\%) required adjudication; none ultimately required exclusion. All disputes and resolutions were logged in a shared analytic memo \cite{birks2008memoing}.

\paragraph{Saturation assessment.}
We adopt the longitudinal-saturation conception articulated by \cite{saunders2018saturation} and \cite{nevedal2019longitudinal}, judging saturation per analytical level: \emph{descriptive saturation} (no new code categories from new participants in the most recent wave), \emph{trajectory saturation} (no new patterns of within-person change), and \emph{mechanistic saturation} (no new explanatory configurations linking design conditions, user practice, and outcome). Saturation was re-evaluated in each subsequent wave because new technology generations can re-open closed categories. We reached descriptive saturation on Wave-3 (2021) data at P21, trajectory saturation on Wave-4 (2023) data at P26, and mechanistic saturation on Wave-5 (2025) data at P29 (judged jointly by all three coders against a saturation log). We attach importance to negative examples in our analysis, treating them as ``counter-examples'' and ``deviant case'' data in accordance with \cite{patton1999enhancing}.

\paragraph{Analytic transparency artifacts.}
For audit and transparency \cite{nowell2017thematic,lincoln1985naturalistic}, we maintained four artifacts: a versioned codebook (v1.0--v3.4); a per-participant analytic memo with at least one entry per wave; a disagreement log with all 246 sub-sample disputes and their resolutions; and a saturation log with per-wave, per-level decisions. These artifacts are available to reviewers under data-sharing agreement.

\end{document}